\documentclass[pre,twocolumn,nofootinbib]{revtex4}
\usepackage[utf8]{inputenc}
\usepackage{amsmath}
\usepackage{amsfonts}
\usepackage{amssymb}
\usepackage{graphicx}
\usepackage{grffile}
\usepackage{color}
\usepackage{multirow}

\begin{document}
\title{ Time Scales of Fickian Diffusion and  the Lifetime of Dynamic Heterogeneity}
\author{Rajsekhar Das$^{1}$}
\email{rajsekhard@tifrh.res.in}
\author{Chandan Dasgupta$^{2,3}$}
\email{cdgupta@iisc.ac.in}
\author{Smarajit Karmakar$^{1}$}
\email{smarajit@tifrh.res.in}
\affiliation{
$^1$ Tata Institute of Fundamental Research, 
36/P, Gopanpally Village, Serilingampally Mandal,Ranga Reddy District, Hyderabad 500107, India.\\
$^2$ Centre for Condensed Matter Theory, Department of Physics,
Indian Institute of Science, Bangalore 560012, India.\\
$^3$ International Centre for Theoretical Sciences, Bangalore 560089, India.
}

\begin{abstract}
Dynamic heterogeneity, believed to be one of the hallmarks of the 
dynamics of supercooled liquids, is expected
to affect the diffusion of the particles comprising the liquid. 
We have carried out extensive molecular dynamics simulations of two model glass 
forming liquids in three dimensions to study the time scales
at which Fick's law of diffusion starts to set in. 
We have identified two different Fickian times cales: one at which the mean squared
displacement begins to show a linear dependence on time, and another one at which the 
distribution of particle displacements becomes Gaussian. These two 
times scales are found to be very different from each other and from the  
$\alpha$ relaxation time in both systems. An interesting 
connection is found between one of these Fickian time scales and the time scale obtained 
from the bond-breakage correlation function. We discuss 
the relation among these different times cales and their connection
to dynamic heterogeneity in the system. 
\end{abstract}

\maketitle
\section{Introduction}
One of the important unsolved problems in condensed
matter physics is to understand the complex dynamics of
supercooled liquids near the glass transition. In last
several decades, immense efforts have been made to understand the rapid rise of the viscosity and the slowing down of
the dynamics of supercooled liquids as the
glass transition is approached~\cite{09Cav,11BB,14KDS,05Berthier,06BBMR,08BBCGV,09KDS,15KDSRoPP,RFOT,RFOT1,RFOT2,Ediger2000}. It is now well-accepted that 
the slow dynamics becomes increasingly heterogenious near the glass transition. Numerous studies have been performed
to understand this behavior~\cite{Ediger2000,RankoRichart2002,Andersen6686}. 
While the origin of dynamic heterogeneity is still not fully understood, the lifetime of 
these heterogeneities is also debated. In an earlier
study~\cite{Szamel2006}, it has been shown that the lifetime of dynamic heterogeneity grows faster than the
$\alpha$-relaxation time, whereas in some other works, this time scale has been shown to 
be comparable to the $\alpha$-relaxation time~\cite{Ediger1995,Ediger1999,
Deschenes255,Bohmer1996,Bohmer1998}.

In recent studies~\cite{Rajsekhar2018, 
KarmakarPrl2016}, it has been shown that the short-time $\beta$-relaxation in 
supercooled liquid is a cooperative process. 
It is also shown that the temperature dependence of the length 
scale associated with the cooperative $\beta$-relaxation process is the 
same as that of the dynamic heterogeneity length scale obtained at the 
$\alpha$-relaxation time. Note that the $\alpha$-relaxation time can be many 
orders of magnitude larger than the $\beta$-relaxation time, especially at 
low temperatures. Thus, the dynamics of a supercooled liquid is  
heterogeneous at time scales that can be as short as the $\beta$-relaxation time, and 
these heterogeneities 
survive up to a time scale that can be much larger than the 
$\alpha$-relaxation time. The difference between the survival time of 
dynamic heterogeneity and the $\alpha$-relaxation time increases with increasing
supercooling. In earlier studies \cite{Berthier1797,Szamel2010,
Lacevic2003, PhysRevLett.104.165703}, it has been suggested that dynamic 
heterogeneity in glass-forming liquids of binary hard-sphere systems can 
survive up to a time scale that is larger than the $\alpha$-relaxation time
by a factor of a few tens. 

It is well-known that the mean squared displacement (MSD), 
$\langle \Delta r^2 \rangle$, computed for a
supercooled liquid shows an initial ballistic growth 
with time and then tends to a plateau at an intermediate time scale. The 
plateau is followed by a diffusive region. The plateau in the MSD appears because of the hindrance of the
motion of a particle caused by the ``cage'' formed by its neighbours. The 
diffusive regime sets in when the particles manage to break out of the 
cage, but the dependence of the MSD on time does not 
become linear until a time scale that is larger than the 
$\alpha$-relaxation time.  Although the MSD shows diffusive behaviour 
at this time scale, the dynamics of the system remain spatially heterogeneous. 
In Ref.~\cite{Szamel2006}, it has been shown that the distribution 
of the displacements of the particles becomes Gaussian at a time scale that 
is at least $30 - 40$ times larger than the $\alpha$-relaxation time. 
True Fickian diffusion starts after this time scale. Thus there 
exists two different time scales, related to diffusion. One of these is the time at which the slope 
of the MSD versus time in a log-log plot becomes unity. We call this time scale 
$\tau_D$. The other one is the time at which the distribution of the displacements
of the particles becomes Gaussian. We call this time
$\tau_F$. A similar time scale, $\tau_H$, is obtained via the Binder cumulant of 
the van Hove correlation function.

It is interesting to inquire whether the Fickian time scales defined above are
related to the lifetime of dynamic heterogeneity. In Ref.~\cite{Szamel2006}, it has been suggested that the 
time scale obtained via the distribution of single particle displacements 
may provide a lower bound for the lifetime of dynamic heterogeneities 
present in the system, but the actual lifetime of the heterogeneities may be much larger. 
The violation of the Stokes-Einstein relation between the diffusion coefficient
and the viscosity~\cite{Hansen1986,Einstein,LANDAU1987227}, another poorly understood phenomenon in glassy dynamics, is
also believed to be closely related to dynamic heterogeneity.

In this paper, we have addressed some of the issues mentioned above
via extensive molecular dynamics simulations of two well known model glass 
forming liquids in three spatial dimensions. The primary objective of this study 
is to estimate the time scales $\tau_D$, $\tau_F$ and 
$\tau_H$ in different generic glass-forming liquids and compare them with
other well known important time scales in the system, such as 
the $\alpha$-relaxation time $\tau_\alpha$ and the bond-breakage time scale $\tau_{BB}$ (to be defined later) 
We have looked for the 
existence of correlations among these different time scales and tried to figure out
their consequence in understanding the dynamics 
of glass forming liquids approaching the glass transition. We find that the time scales
of diffusion, $\tau_D$, $\tau_F$ and $\tau_H$, are longer than $\tau_\alpha$ in the
temperature range considered here. The temperature dependence of $\tau_F$ is similar
to that of  $\tau_H$ and their growth with decreasing
temperature is faster than that of $\tau_\alpha$. The growth of $\tau_D$ as the 
temperature is reduced is, on the other hand, slower than that of $\tau_\alpha$. We
find that the relations between pairs of these time scales are described by power
laws at low temperatures, so that the glass transition temperatures obtained from
Vogel-Fulcher-Tammann~\cite{Fulcher1925,Vogel1921}(VFT) or mode
 coupling~\cite{gotze1991liquids,Gotze_1992} (power law) fits to the temperature dependence of these time scales
are the same within error bars. We also find an intriguing connection between $\tau_D$
and the bond-breakage time scale $\tau_{BB}$. The implications of these findings for the 
violation of the Stokes-Einstein relation and the kinetic fragility are discussed.

The rest of the paper 
is organized as follows. We present the details of the models studied and the simulation 
methodology in the Model $\&$ Method section. We then introduce 
different correlation functions that are computed to estimate
different time scales and present our results for two different generic 
glass-forming liquids. Finally we conclude with a discussion of the importance of 
these different time scales for understanding the intricate role played by 
dynamic heterogeneity in the dynamics of glass-forming liquids. 

\section{Models \& Methods}
We have studied two different model glass forming liquids in three 
dimensions.
\begin{itemize}
\item{\bf 3dKA Model\cite{95KA} :} This is a well known $80:20$ binary Lennard-Jones 
mixture in three dimensions. Particles interact via the following
pairwise interaction 
\begin{equation}
V_{\alpha\beta} (r)= 4\epsilon_{\alpha\beta}\left[\left(\frac{\sigma_{\alpha\beta}}{r}\right)^{12}- \left(\frac{\sigma_{\alpha\beta}}{r}\right)^{6}\right],
\end{equation}
where $\alpha,\beta \in \{A,B\}$  and $\epsilon_{AA}=1.0$, $\epsilon_{AB}=1.5$,
$\epsilon_{BB}=0.5$, $\sigma_{AA}=1.0$, $\sigma_{AB}=0.80$,
$\sigma_{BB}=0.88$. We cut off the interaction potential at 
$2.50\sigma_{\alpha\beta}$. A quadratic polynomial is used to make the 
potential and its first two derivatives smooth at cutoff distance. The temperature 
range for the simulations is $T \in [0.450 - 1.000]$ and the system size is
$N = 8000$.
 
\item{\bf 3dR10 Model~\cite{12KLP} :} It is  a $50:50$ binary mixture where the 
particles interacts via
\begin{equation}
V_{\alpha\beta}(r) = \epsilon_{\alpha\beta}\left(\frac{\sigma_{\alpha\beta}}{r}\right)^n,
\end{equation}
with $n= 10$. The potential is cut off at a distance $1.38\sigma_{\alpha\beta}$ 
and again a quadratic polynomial is used to make potential and its first 
two derivative smooth at the cutoff distance. Here $\epsilon_{\alpha\beta} = 
1.0$, $\sigma_{AA} = 1.0$, $\sigma_{AB} = 1.22$ and $\sigma_{BB} = 1.40$. 
The temperature range for the simulations is $T \in [0.520-1.000]$ and $N = 10000$.

For all the studied models we have performed NVT molecular dynamics (MD) simulations
with a modified leap-frog algorithm. We use the Berendsen thermostat to keep the temperature constant. Our integration time step is $dt = 0.005$. The number of MD steps are $10^7-10^8$, depending on the temperature.
\end{itemize}

\section{Correlation functions}
\label{correlationFunc}

\begin{figure}
\begin{center}
%\vspace{-1cm} 
\includegraphics[scale=0.34]{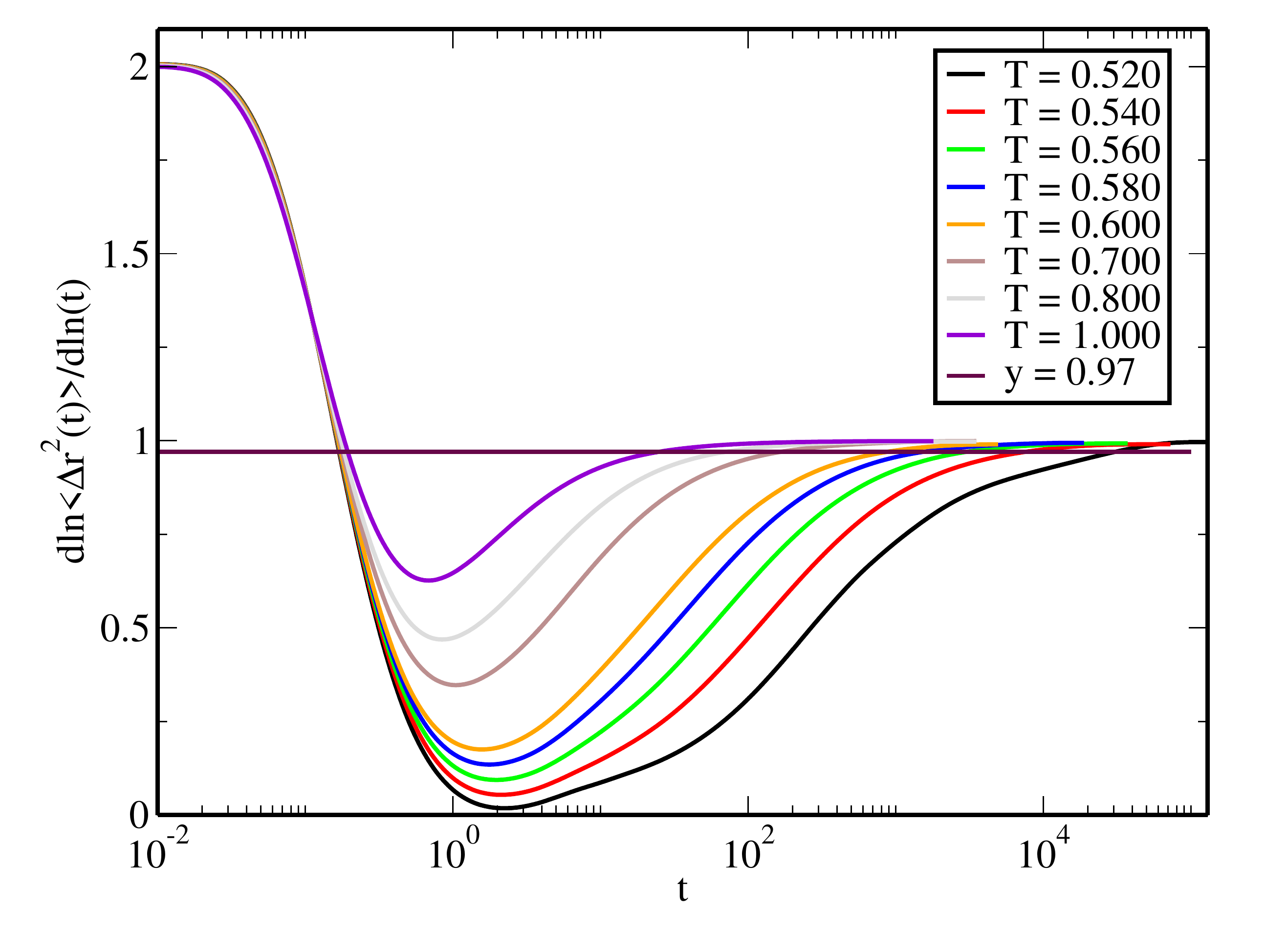}%\quad
\hskip 0.1 cm
\includegraphics[scale=0.38]{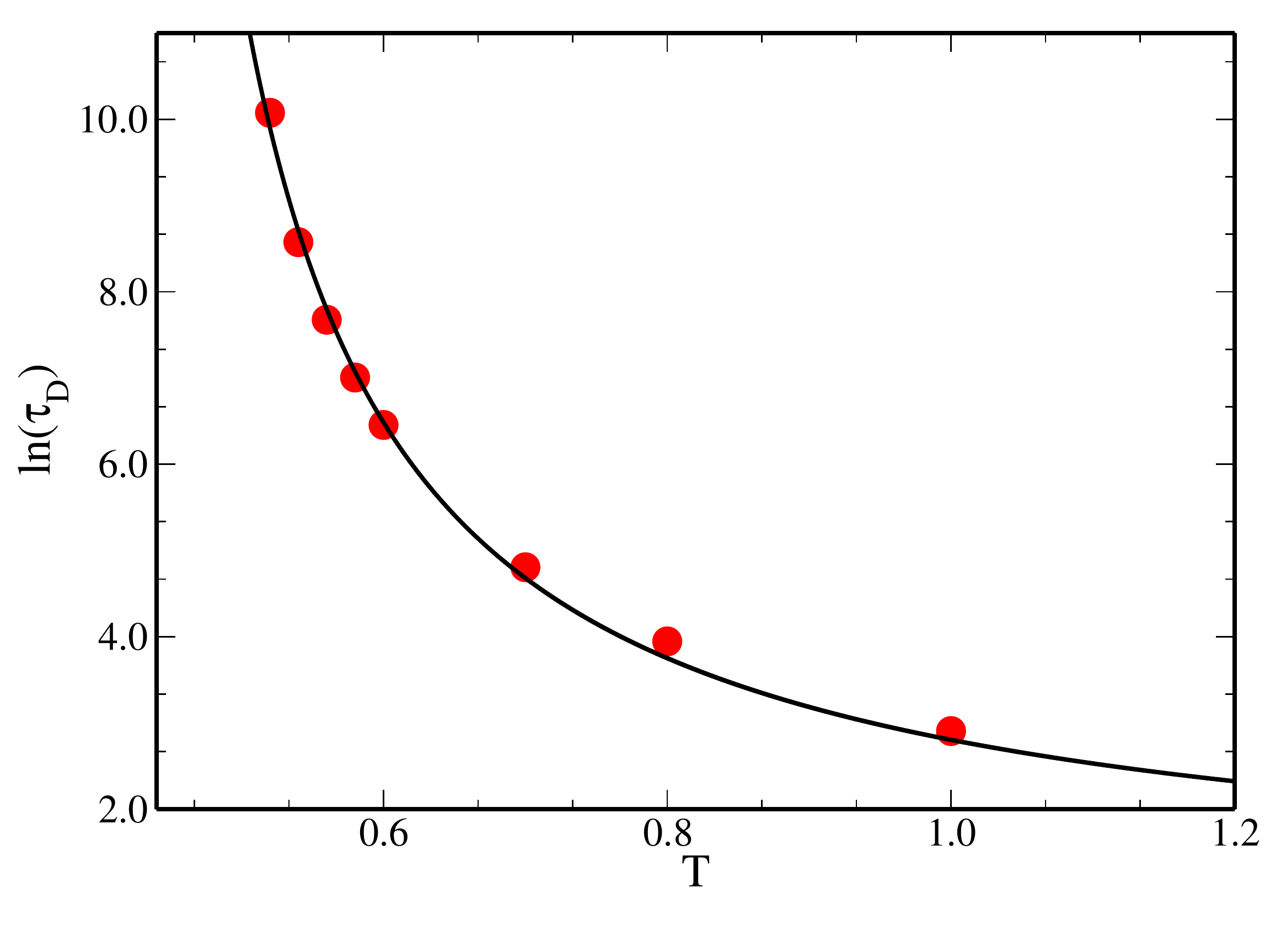}
\caption{\textbf{Top panel:} Log-derivative of MSD is plotted versus time for the 3dR10 model. The horizontal line corresponds to $ 0.97$. \textbf{Bottom Panel:} The dependence of the logarithm of the time scale $\tau_D$  on the temperature. The line passing through the data points is a fit of the data to the VFT form, $\tau_\alpha = \tau_0\exp\left[\frac{A}{T-T_{VFT}}\right]$.}
\label{3dKA_logrdist}
\end{center}
\end{figure}

\subsection{Overlap Correlation function}

We characterize the dynamics by calculating the well known two-point 
density-density correlation function $Q(t)$. It measures the overlap 
between two configurations separated by time $t$. The self part of 
this correlation function is defined as
\begin{equation}
Q_s(t) = \frac{1}{N}\left[\left<\sum_{i=1}^{N}w
\left( |\vec{r}_i(t) - \vec{r}_i(0)|\right)\right>\right],
\end{equation}
where $w(x)$ is a window function with 
\begin{equation}
w(x)=
\begin{cases}
1.0 & \text{if $x \le a$}\\
0 & \text{otherwise.}
\end{cases}
\end{equation}
The value of $a$ is chosen to be $0.30$ which is close 
to the value at which the MSD forms a plateau. The window function  is chosen to remove the possible 
decorrelation due to the vibrational motion of the 
particles inside their cage.  A different choice of the parameter $a$ does
not change the results qualitatively.

\subsection{van Hove Correlation function}
\label{vanHovecorr}
The van Hove correlation function measures the probability that 
a particle has displacement $x$ after time $t$.
The self part of the van Hove correlation function~\cite{vanHove1954} 
is defined as
\begin{equation}
G_s(x,t) =\sum_{i=1}^{N}\delta(x-x_i(t)+x_i(0)),
\end{equation}
where $x_i(t)$ is the $x$-component of the position vector of particle
$i$ at time $t$.
\begin{figure}[!htpb]
\begin{center}
\includegraphics[scale=0.37]{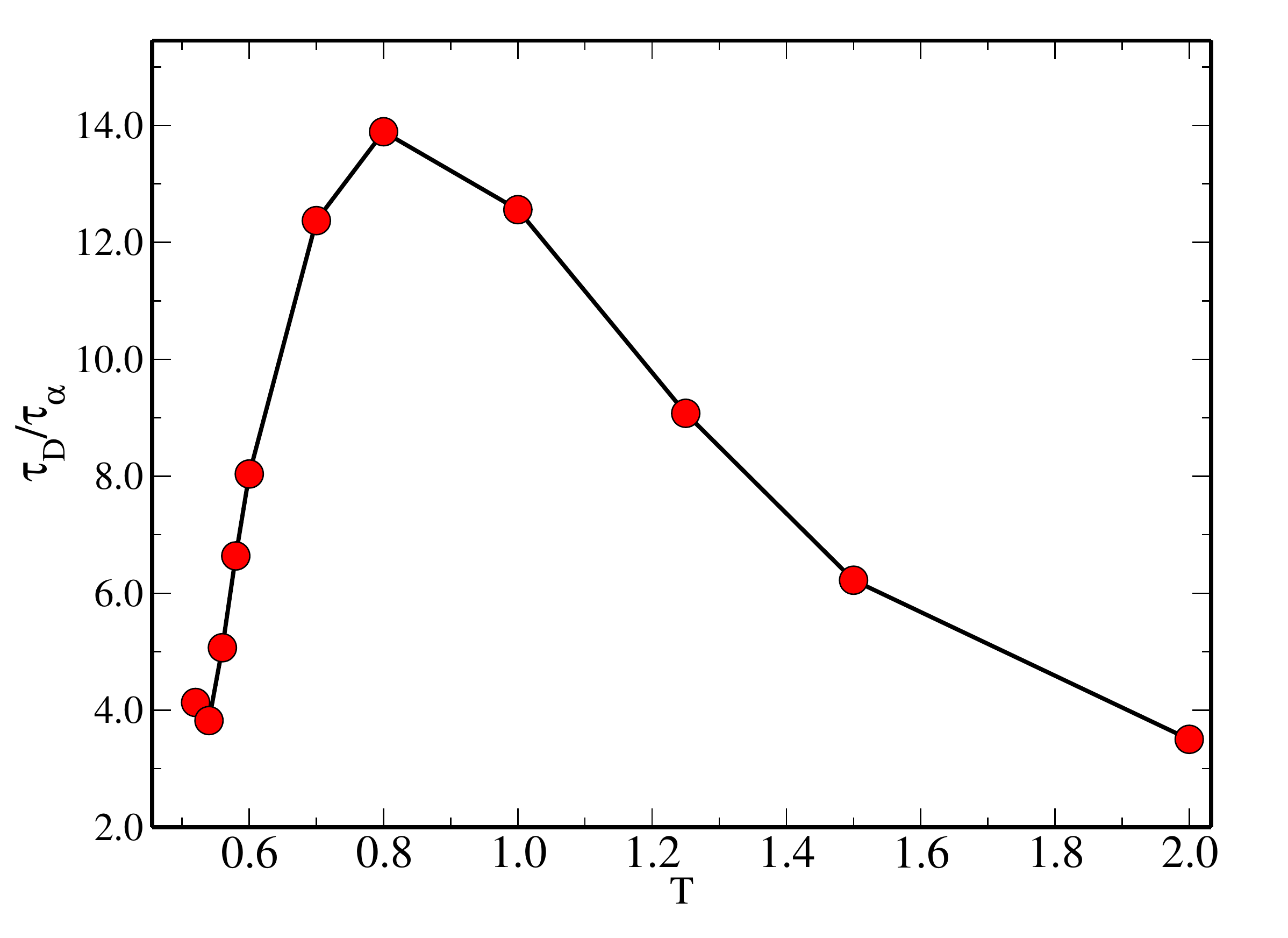}
%\hskip +0.2cm 
\includegraphics[scale=0.35]{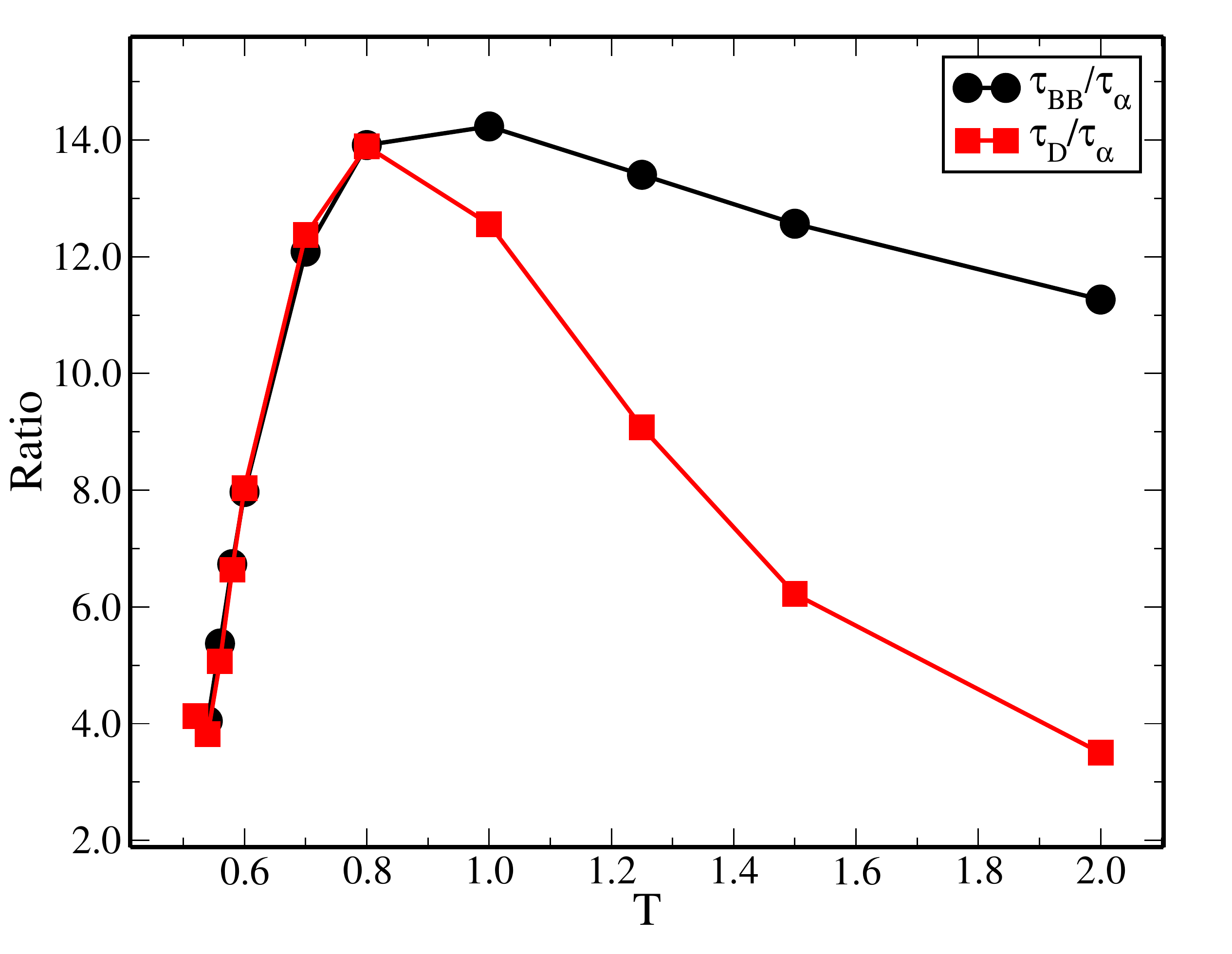}
\caption{\textbf{Top panel:} Ratio of the time scales $\tau_D$ and $\tau_\alpha$ for the 3dR10 model.  \textbf{Bottom Panel:} Comparison of this ratio to the ratio of $\tau_{bb}$ and $\tau_\alpha$ for the same model.}
\label{ratio_tau_msd1_tau_alpha}
\end{center}
\end{figure}
\subsection{Bond-Breakage Correlation Function}
The bond-breakage correlation function is defined in the following 
way\cite{Shiba2016,Yamamoto1997,Indra2018}. At $t = t_0$ a 
pair of particle $i$ (of type $\alpha$) and $j$ (of type $\beta$) is considered to be bonded if
\begin{equation}
r_{ij}(t_0) = |\vec{r}_i(t_0) - \vec{r}_j(t_0)| \le \sigma_{\alpha\beta}.
\end{equation}
If $r_{ij}(t) \le 1.35\sigma_{\alpha\beta}$, the bond 
is said to have survived at time $t$. We calculate number of bonds that
survive at time $t$ and the bond-breakage correlation function is defined 
as the ratio of this number to the initial number of bonds.
The bond-breakage time scale $\tau_{BB}$ is the time at which the correlation 
function goes to $1/e$ of its initial value.

\section{Results}

\subsection{Time Scale from the Mean Square Displacement}
As mentioned before, the mean square displacement (MSD) shows three distinct regimes 
for a supercooled liquid: initial ballistic growth {\it i.e.} 
$\langle\Delta r^2 \rangle \sim t^2$, then a plateau followed by a 
diffusive regime {\it i.e.} $\langle\Delta r^2\rangle \sim t$. So a
plot of the logarithmic derivative of the MSD with respect to time starts 
from $2$, goes to a minimum and then eventually goes to 
$1$, corresponding to the above three distinct regimes. The time scale 
where it goes to $1$ is of importance as this is the time at which
the system begins to show diffusive behavior. We have calculated this 
time scale, $\tau_D$, for different temperatures for both the model
systems. In the top panel of Fig.~\ref{3dKA_logrdist}, we show the logarithmic 
derivative of the MSD as a function of time for all the studied temperatures
for the 3dKA model. We have taken $\tfrac{d\ln{\langle\Delta r^2 \rangle}}{d\ln t} = 0.97$ 
to define the time scale $\tau_D$. The temperature dependence of this time scale 
is shown in the bottom panel of the same figure. The temperature dependence can be fitted very well using the VFT form with the divergence
temperature $T_{VFT} \simeq 0.29$.

To understand the mutual relationship between $\tau_D$ and $\tau_\alpha$, 
we have plotted the ratio of these two time scales ($\tau_D/\tau_\alpha$) as 
a function of temperature in the top panel of Fig.~\ref{ratio_tau_msd1_tau_alpha}. 
As the temperature is decreased, this ratio first increases, reaches a maximum near $T=0.8$,
and then decreases as the temperature is lowered further. The temperature at which $\tau_D/\tau_\alpha$ peaks is
close to the onset temperature $T_{on}$\cite{Sastry:1998aa}  for this model. Thus, at 
temperatures lower than $T_{on}$  which corresponds to the temperature below which the effects 
of the potential energy landscape
becomes important, the diffusion time
scale $\tau_D$ grows more slowly than the structural relaxation time $\tau_\alpha$ with decreasing temperature.  

We have also calculated the bond-breakage time scale $\tau_{BB}$ following the 
method described in Sec.~\ref{correlationFunc}. It is the time scale 
where the most of the particles have gone though some amount of reshuffling
of their neighbors, as measured using the breakage of nearest neighbor bonds.
The temperature dependence of $\tau_{BB}/\tau_\alpha$  is similar to that of
$\tau_D/\tau_\alpha$, as shown in the bottom panel of Fig.~\ref{ratio_tau_msd1_tau_alpha}. 
In particular, these two ratios exhibit nearly identical temperature dependence for values of $T$ lower than that at which they exhibit a peak. This observation suggests an intriguing connection between diffusion and bond breakage at remperatures lower than
the onset temperature. However, we do not have a clear understanding 
of this behavior. 

\begin{figure}
\begin{center}
\includegraphics[scale=0.33]{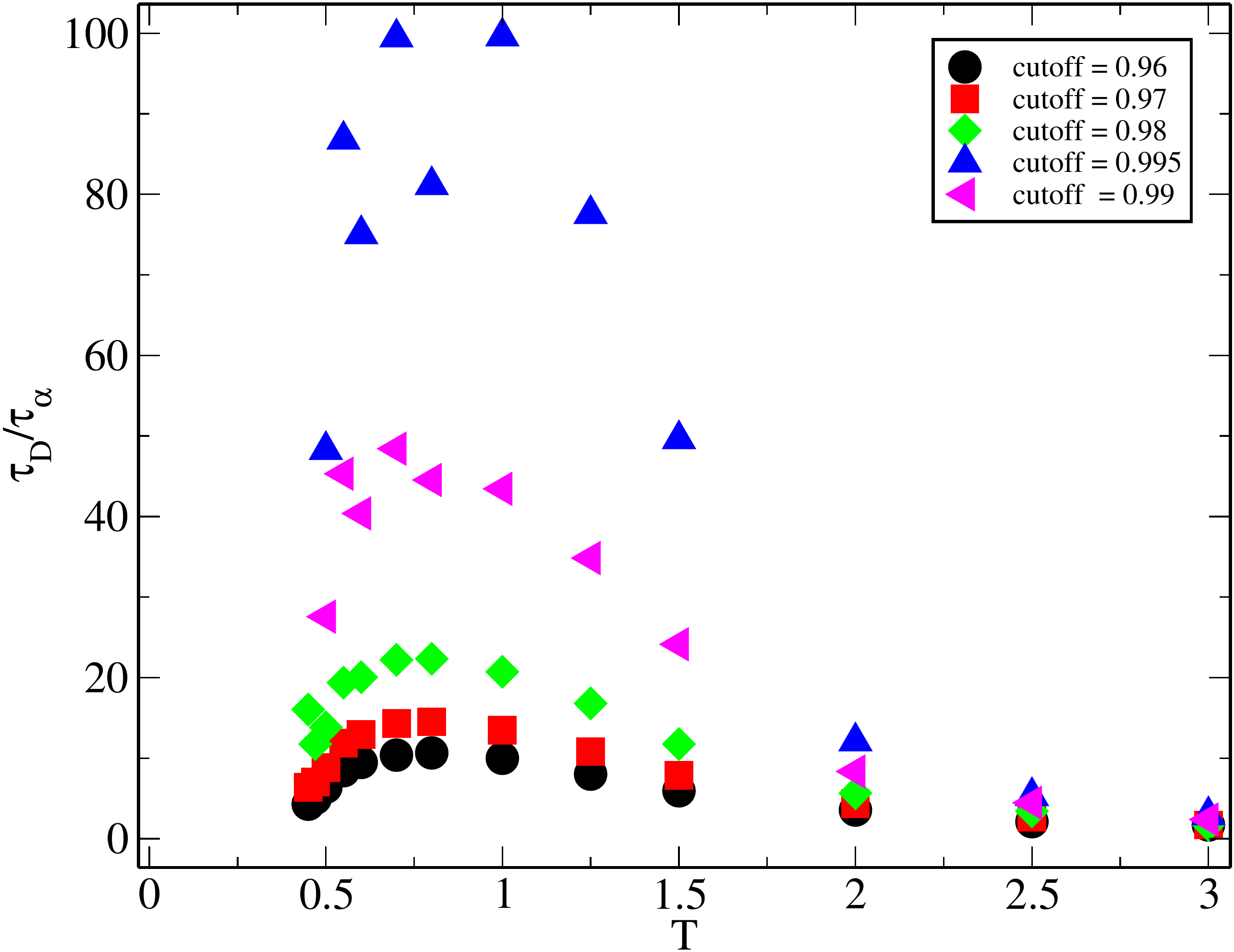}
\caption{Cutoff dependence of $\tau_D/\tau_\alpha$ for the 3dKA model. 
It is clear that that although the value of this ratio depends on the cutoff,  the non-monotonic 
temperature dependence is present for all values of the cutoff. }
\label{cutoffDependenetRatio}
\end{center}
\end{figure}
To calculate $\tau_D$, we have chosen the cutoff value of the logarithmic derivative of the MSD to be $0.97$. Ideally, $\tau_D$ should be calculated where this value actually goes 
to  unity. This is quite challenging as for low temperatures, 
the derivative does not reach the value of unity in our simulation time scale.  Even if the derivative goes to unity within the simulation time scale, the noise in its measurement makes it difficult to  determine accurately the time at which it first reaches this value. To check whether the non-monotonic behavior of the temperature 
dependence of $\tau_D/\tau_\alpha$  depends on how we calculate $\tau_D$, 
we have done a systematic study of this ratio by changing the cutoff. 
Fig.~\ref{cutoffDependenetRatio} shows the cutoff dependent ratio 
for the 3dKA model systems. From the plot, it is clear that the non-monotonic 
behavior is a feature that is independent of the choice of the
cutoff: the non-monotonicity increases systematically as the
cutoff approaches unity.  
\begin{figure}%[h!]
\begin{center}
\includegraphics[scale=0.34]{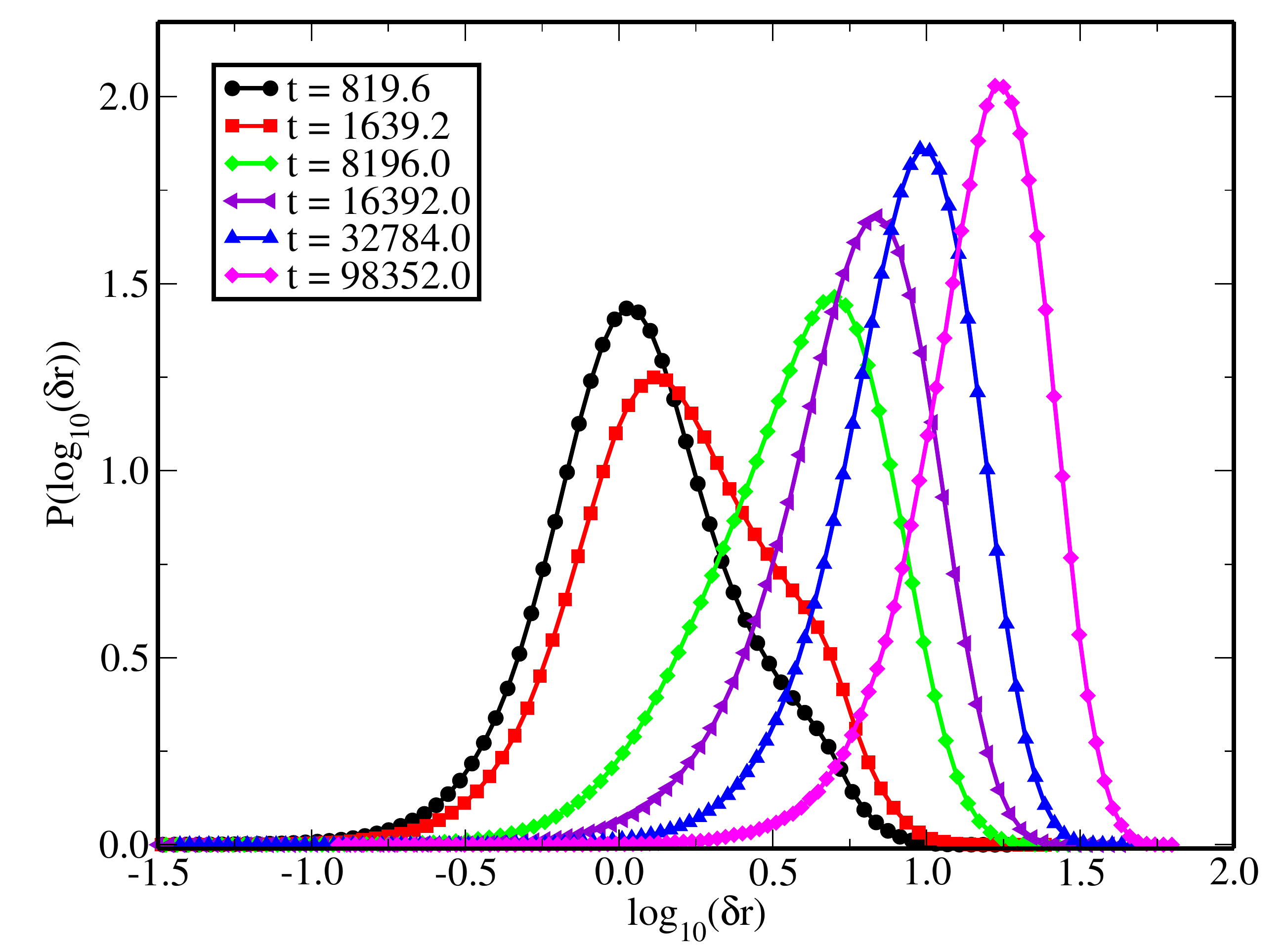}%\quad
\hskip +0.2cm
\includegraphics[scale=0.39]{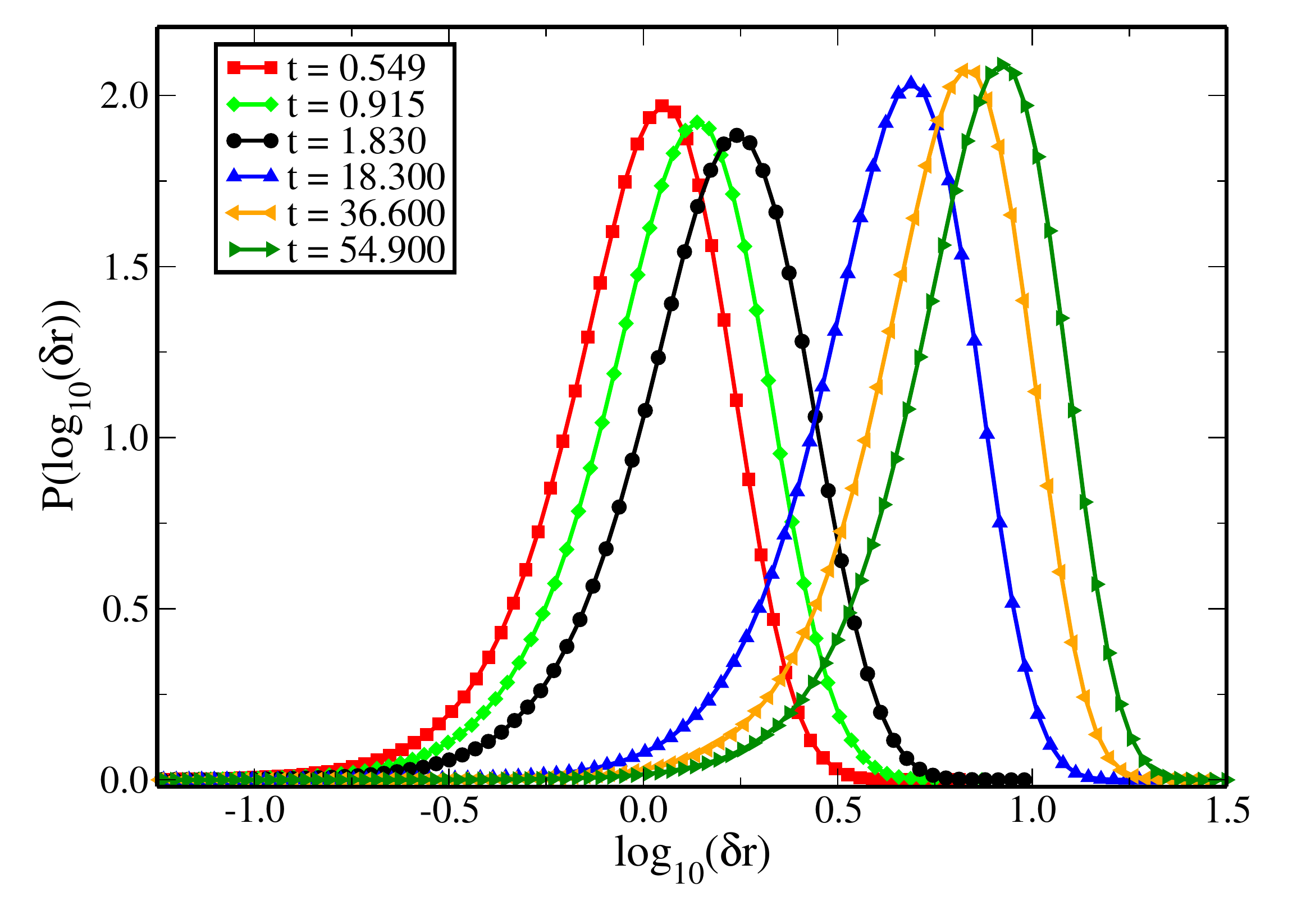}
\caption{\textbf{Top panel:} Distribution of the logarithm of single particle displacements at different times at $T = 0.450$ for the 3dKA model. For relatively short times, the distribution shows signs of bimodality, whereas at longer times it tends to be more Gaussian. \textbf{Bottom Panel:} Similar plot at $T = 1.000$.}
\label{3dKA_SingleParticle}
\end{center}
\end{figure}

\subsection{Time Scale from the Distribution of Single-particle Displacements}
%$\mbox{\Large $\tau_F$}$}

\begin{figure*}
\begin{center}
\hskip -0.15in
\includegraphics[scale=0.35]{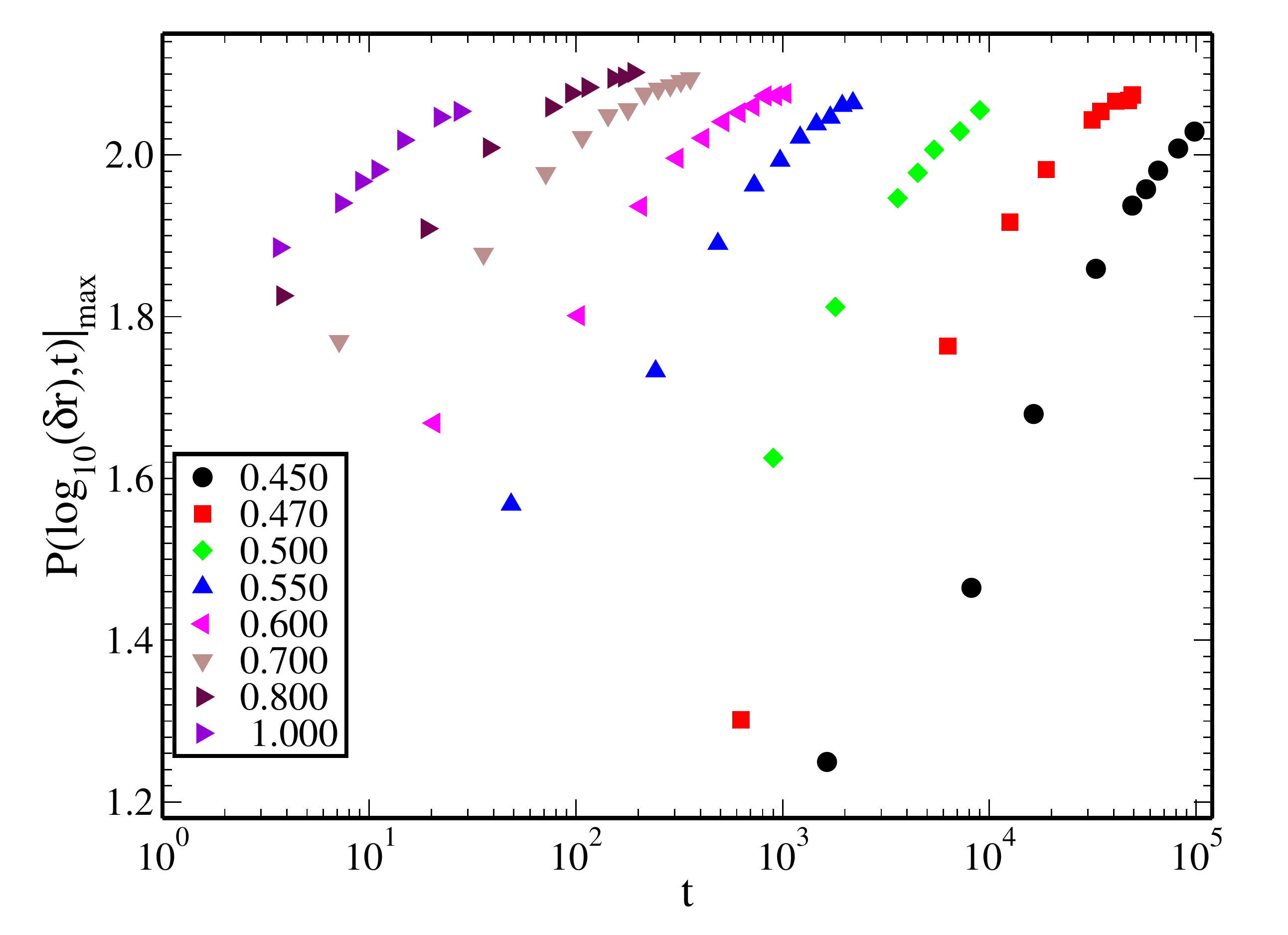}
\hskip -0.25in
\includegraphics[scale=0.365]{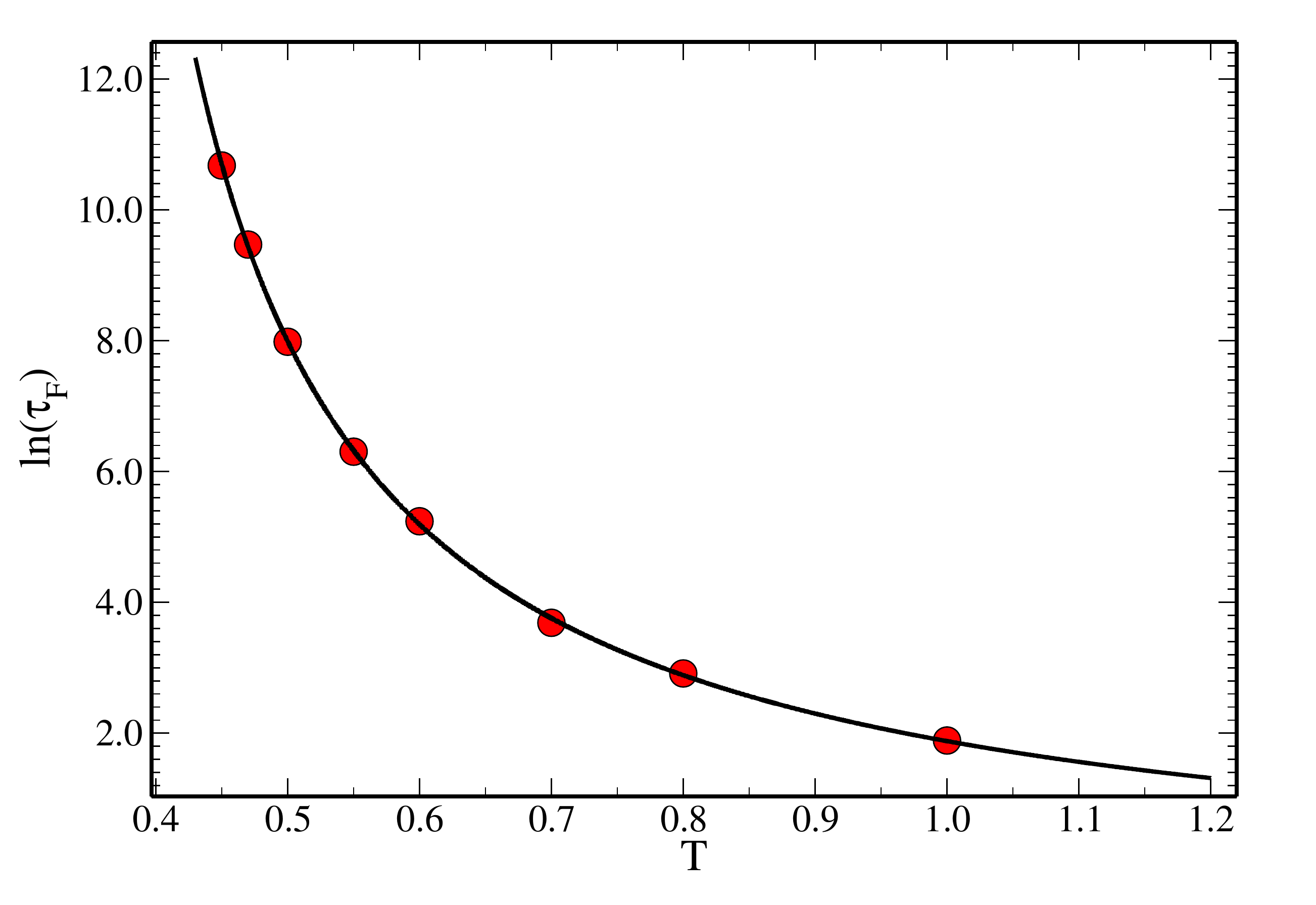}
\caption{\textbf{Left panel:} Value of the maximum of $P(\log_{10}(\Delta r),t)$ is plotted 
against time $t$ for different temperatures for the 3dKA model. The time where it 
reaches $1.92$ is defined as the Fickian time scale $\tau_F$. \textbf{Right Panel:} 
Variation of the logarithm of the time scale $\tau_F$ with temperature $T$. The line passing through the data points is a fit to the VFT form.}
\label{3dKA_taufcalc}
\end{center}
\end{figure*}
Next, we estimate time at the onset of Fickian diffusion by following the 
procedure discussed in~\cite{Szamel2006}. To study the  
onset of Fickian diffusion, it is important to calculate the distribution 
of single particle displacements, $P(\log_{10}(\delta r), t)$ at time $t$. 
As discussed in~\cite{Szamel2006}, this distribution is connected to the self 
part of the van Hove correlation function as
\begin{equation}
P(\log_{10}(\delta r),t) = \ln{10}\, 4\pi \delta r^3G_s(\delta r,t)\ .
\end{equation} 
\begin{figure}[!h]
\begin{center}
\hskip -0.3in
\includegraphics[scale=0.39]{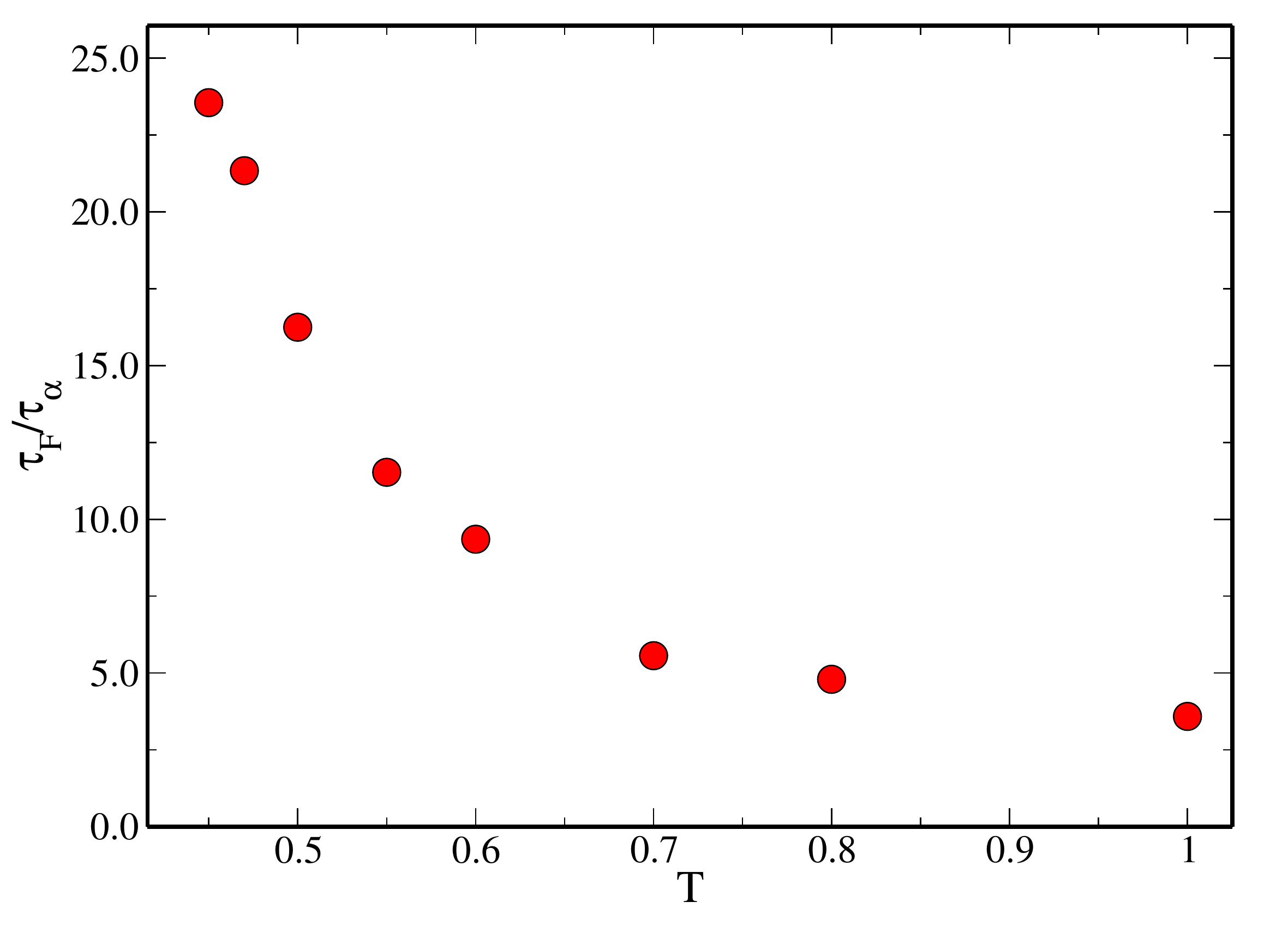}
\caption{The ratio of $\tau_F/\tau_\alpha$ is plotted as a function of $T$. 
Note that $\tau_F$ increases much faster than $\tau_\alpha$
with decreasing temperature.}
\label{3dKA_taufcalc_ratio}
\end{center}
\end{figure}

\begin{figure}
\begin{center}
\includegraphics[scale=0.26]{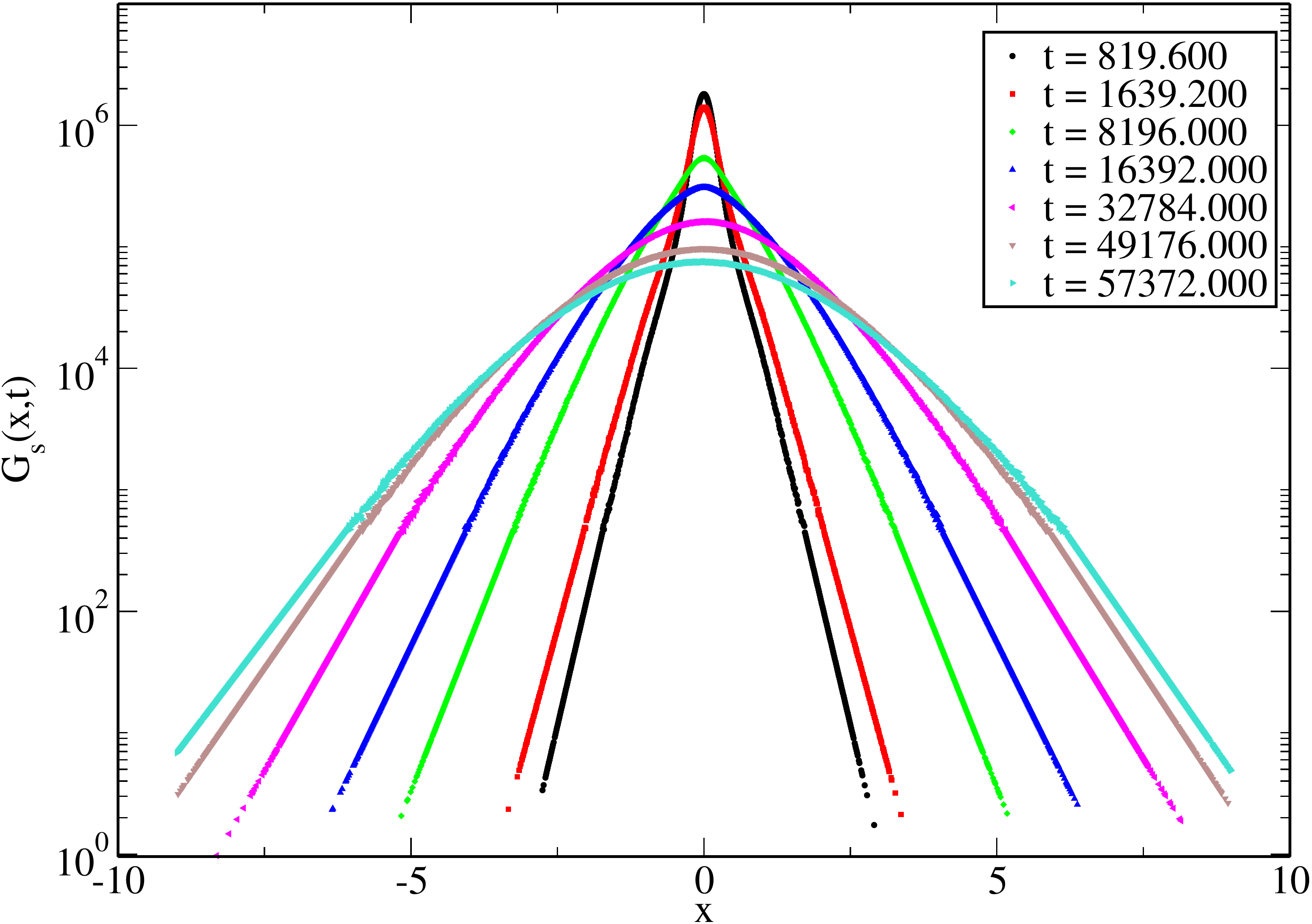}\quad
\includegraphics[scale=0.26]{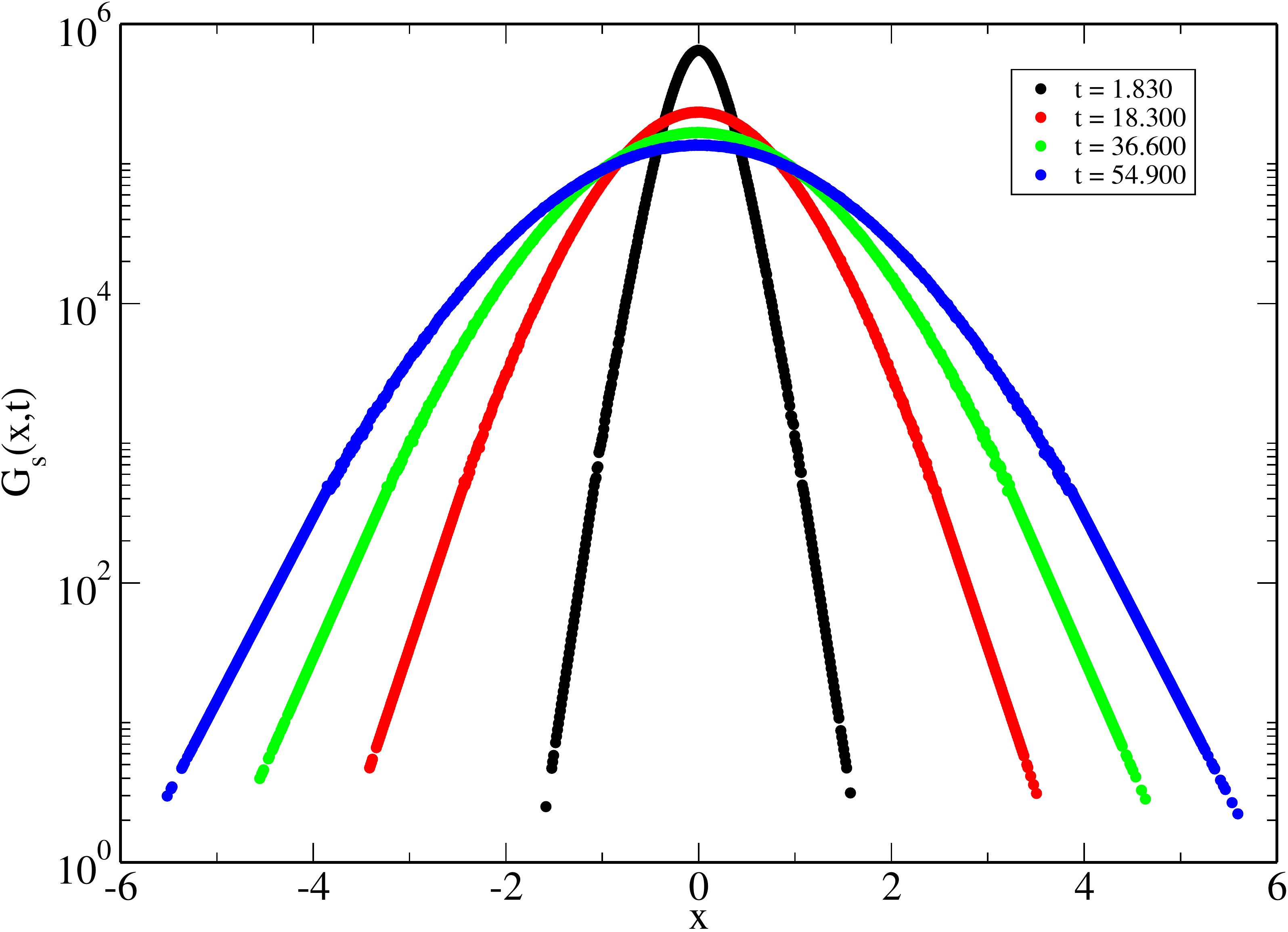}
\caption{{\bf Top panel:} The van Hove correlation function at different 
times at $T = 0.450$ for the 3dKA model. {\bf Botom Panel:} Similar plot 
for $T = 1.000$. }
\label{VanHove_dist}
\end{center}
\end{figure}

\begin{figure}[!h]
\begin{center}
\includegraphics[scale=0.4]{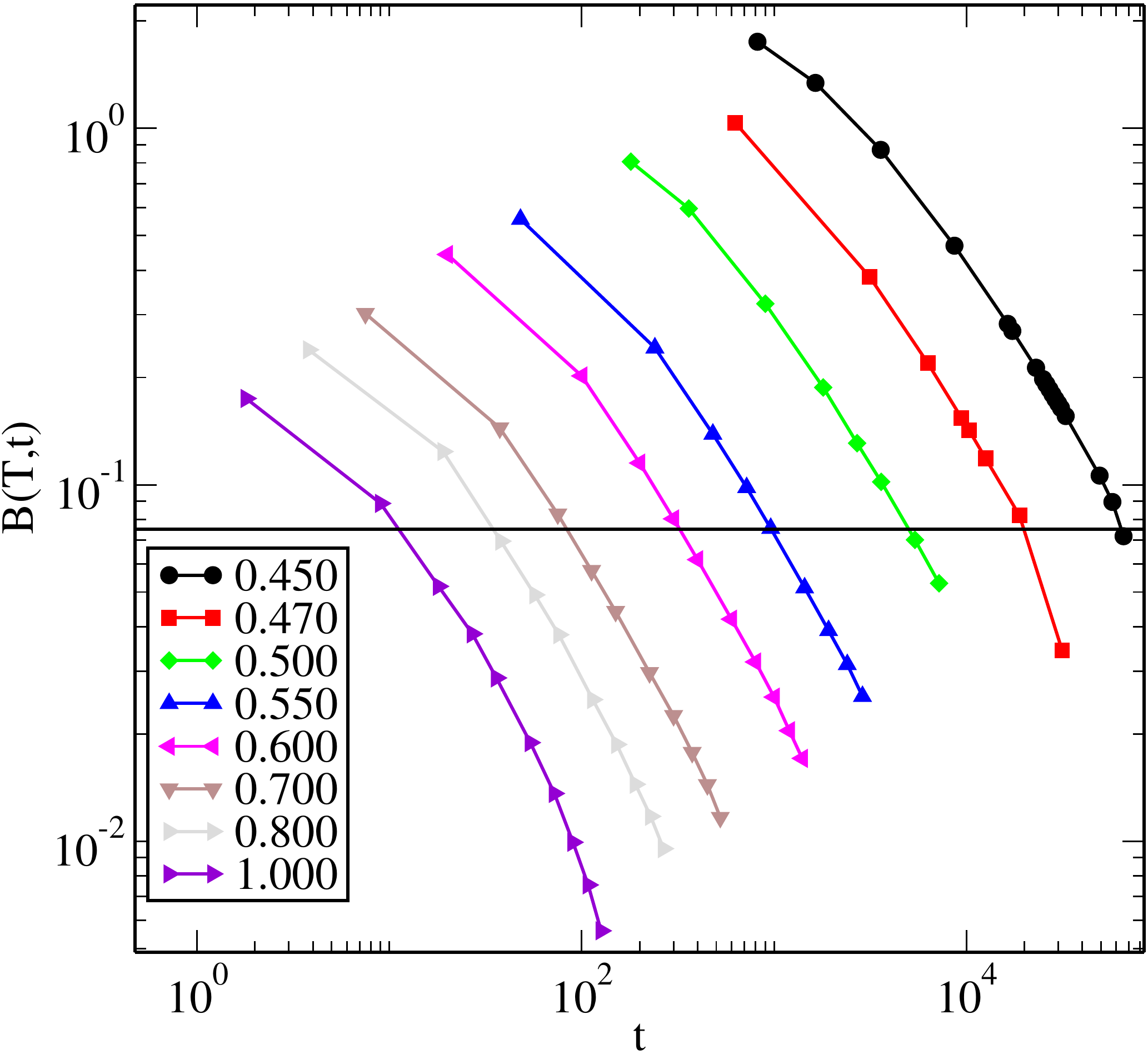}\hskip +0.2in
\vskip +0.1in
\includegraphics[scale=0.44]{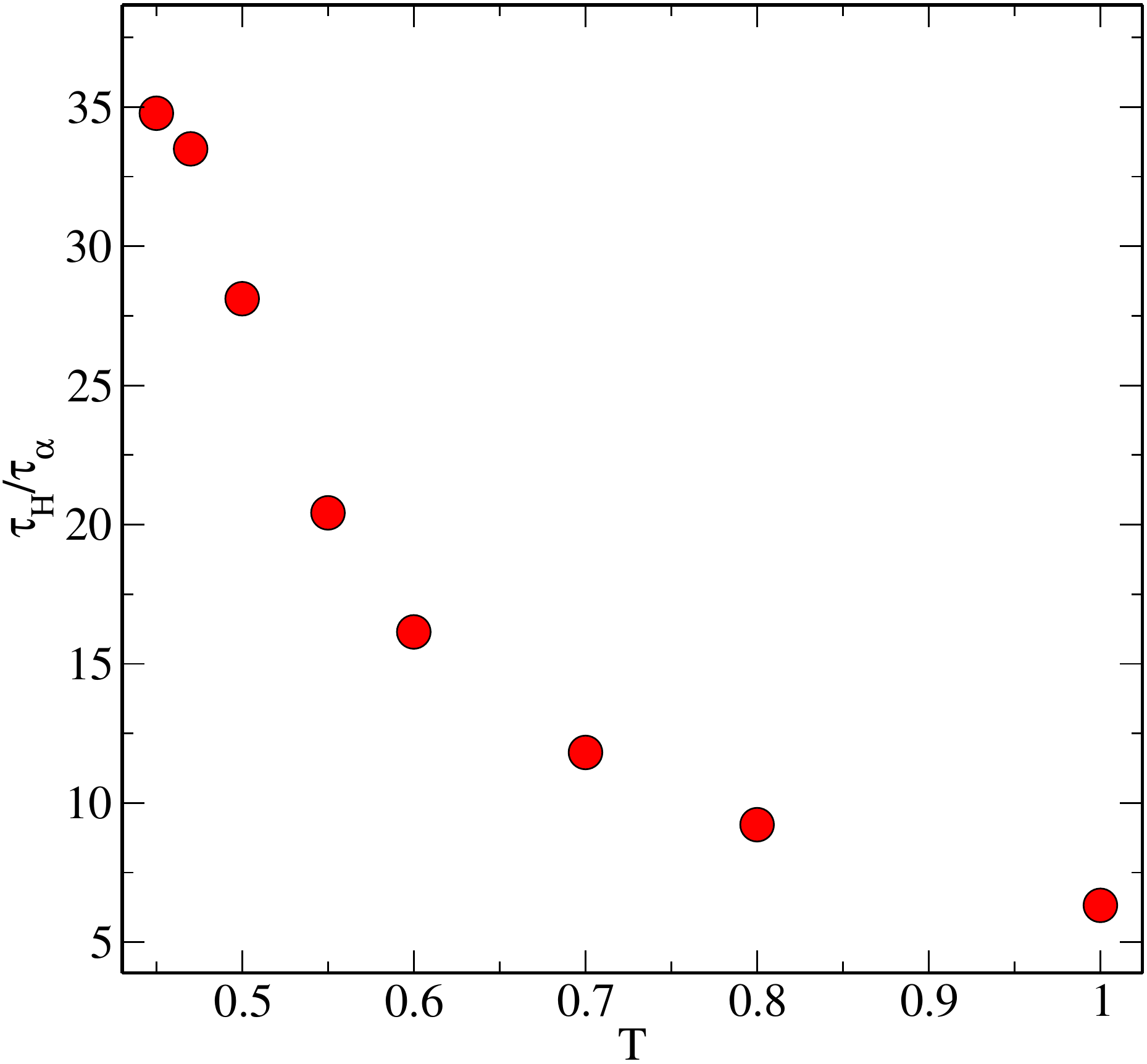}
\caption{\textbf{Top panel:} The Binder Cumulant of the van Hove correlation function for different times and all studied temperatures. The horizontal line corresponds to  $B(T,t) = 0.07$. The time scale $\tau_H$ is defined as the time at which $B(T,t)$ attains this value. \textbf{Botoom panel:} The temperature dependence of the ratio $\tau_H/\tau_\alpha$ .}
\label{3dKA_binderCumulant}
\end{center}
\end{figure}

If the motion of the particles is governed by Fick's law of 
diffusion, the distribution of single particle displacements i.e., 
$P(\log_{10}(\delta r,t))$, should become Gaussian and remain so at longer times. Moreover, the peak value of the distribution 
should reach a value $\approx 2.13$. Any deviation from this behavior would indicate
non-Fickian diffusion as well as dynamic heterogeneity~\cite{Miyazaki_2011,Szamel2006}.
Fig.~\ref{3dKA_SingleParticle} shows plots for $P(\log_{10}(\delta r,t))$ 
at different times for two different temperatures $T = 0.450$ and $1.000$ for the 3dKA model. 
For the lower temperature ($T = 0.450$), deviations from Gaussian behavior are clearly visible  at smaller times, whereas  the distribution approaches a Gaussian shape for longer times and the peak value of the distribution 
also approaches $\approx 2.13$.

At the higher temperature, $T =1.000$, deviations from the Gaussian shape are not 
very clear, but the peak value reaches the value $\sim 2.13$ only
after a few $\tau_\alpha$. This suggests that even at high temperatures, the
dynamics is affected by the presence of spatial heterogeneity 
at time scales larger than $\tau_\alpha$. We define the time scale of onset of Fickian 
diffusion as the time at which the peak value of the distribution goes to 
$\approx 1.92$, similar to the criterion adopted in Ref.\cite{Szamel2006} 
and refer to this time scale as $\tau_F$. In the left panel of 
Fig.~\ref{3dKA_taufcalc}, we show plots of the peak value of  
$P(\log_{10}(\delta r),t)$ with increasing time for several temperatures. 
In the right panel we show the temperature dependence of the time scale $\tau_F$. The temperature dependence of this time scale can be 
fitted very well to the VFT formula. The line passing through the data points in the right panel shows the VFT fit with $T_{VFT} \simeq0.289$. 
To see the mutual relationship between $\tau_\alpha$ and $\tau_F$,
in Fig.~\ref{3dKA_taufcalc_ratio} we have plotted the ratio ($\tau_F/\tau_\alpha$) for 
all the studied temperatures. Form the plot, one can see that 
$\tau_F$ increases much more rapidly than $\tau_\alpha$ when the temperature 
is decreased. The ratio changes from $ 5$ to $25$ in the
temperature window $T \in [1.000-0.450]$ as shown in 
Fig.~\ref{3dKA_taufcalc_ratio}.

Ideally, one should use a cutoff at
$2.13$ in order to calculate $\tau_F$. We could not do that because reaching 
this value within our simulation time scale is not possible. We have,
however, checked that the difference between $\tau_F$ and the other time scale
$\tau_D$ related to diffusion is not an artifact of using a cutoff lower than $2.13$
in the calculation of $\tau_F$ and a cutoff lower than $1.0$ in the calculation of
$\tau_D$. It is evident from the data that $\tau_F$ is larger than $\tau_D$ and it
increases faster than $\tau_D$
as the temperature is reduced. There are indications that these two time scales 
approach each other at temperatures higher than $3.0$.

\begin{figure}[!h]
\begin{center}
\includegraphics[scale=0.37]{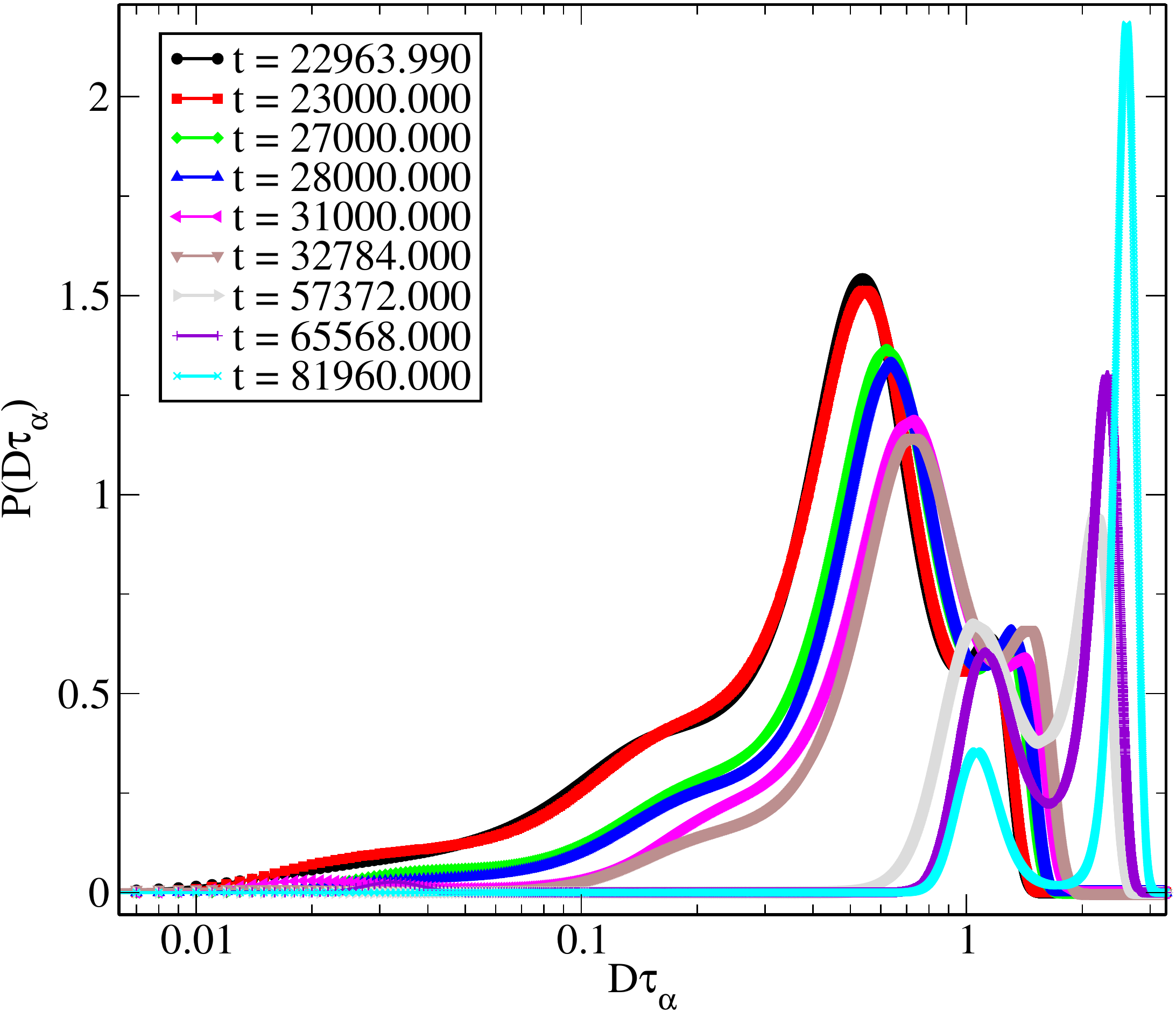}\quad
\includegraphics[scale=0.46]{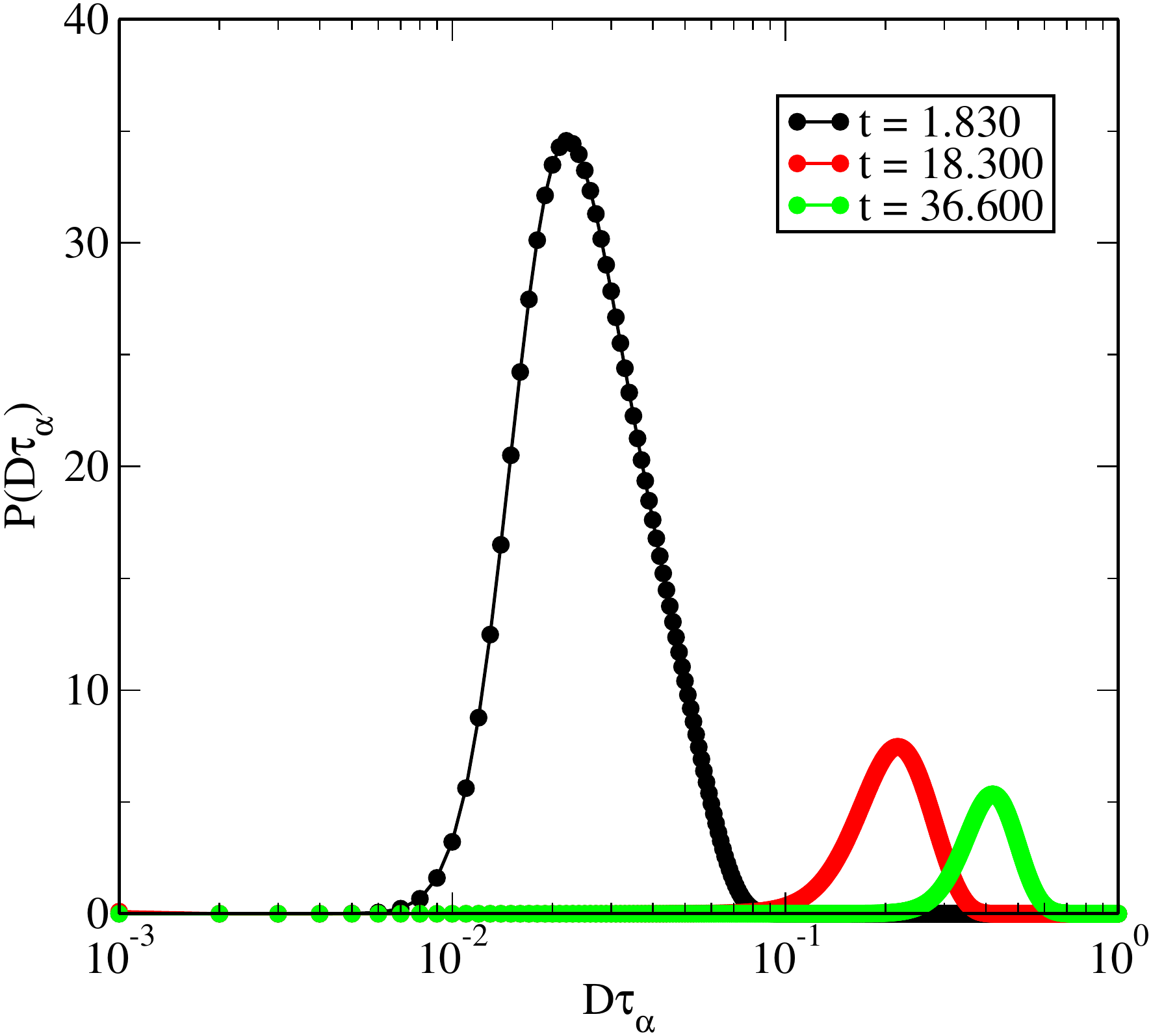}
\caption{{\bf Top panel:} Distribution of diffusion constants for $T = 0.450$ for 3dKA model. {\bf Bottom Panel:} Similar plot for $T= 1.000$ }
\label{Diff_dist}
\end{center}
\end{figure}

\begin{figure}[!h]
\begin{center}
\hskip -0.3in
\includegraphics[scale=0.35]{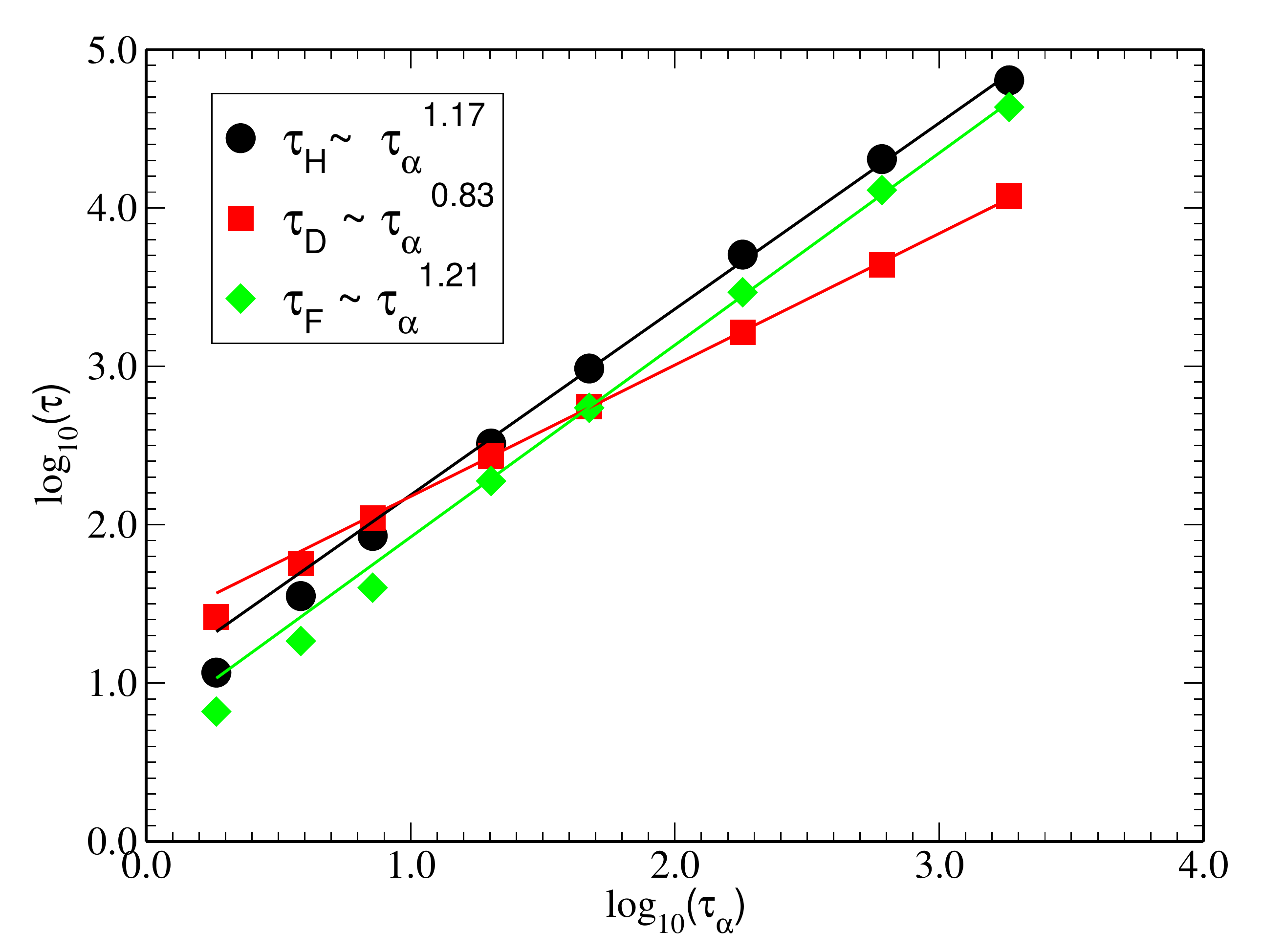}
\caption{Different relaxation times, $\tau_D$, $\tau_F$ and $\tau_H$, are plotted against
the $\alpha$-relaxation time $\tau_\alpha$. Reasonable linear dependence
in a log-log plot suggests that they are related
to one another via power laws. The lines passing 
through the data points are best fits to power laws.}
\label{crossPlot}
\end{center}
\end{figure}

The van Hove function is expected to have a Gaussian form for Fickian diffusion.
In the top panel of Fig~\ref{VanHove_dist}, we have plotted the van 
Hove function for the 3dKA model at $T = 0.450$ for different times. From the 
plots, one can see evidence for non-Gaussian behavior for smaller times and approach
to a Gaussian form for longer times. The bottom panel shows 
similar plots for $T = 1.000$. To get a second measure of the 
degree of non-Gaussianity, we have calculated  the Binder 
cumulant of the van Hove function for different times.
The Binder cumulant, which provides a measure of the excess kurtosis of the distribution, is defined as
\begin{equation}
B(T,t) = 1- \frac{\left\langle x^4\right\rangle}{3\left\langle x^2 \right\rangle^2}.
\end{equation}

If the van Hove function is perfectly Gaussian, 
the Binder cumulant should be zero. For that to happen, one 
needs to run the simulations for very long times. Here we have defined a time scale $\tau_H$
as the time where the Binder cumulant 
of the van Hove function becomes $\approx 0.07$. In the top panel 
of Fig.~\ref{3dKA_binderCumulant}, we show the Binder cumulant for all studied temperatures 
at different times for the 3dKA model. The horizontal
line in the figure corresponds to the value of $0.07$. The 
bottom panel shows the ratio of $\tau_H$ and $\tau_\alpha$ as a function of temperature. 
The temperature dependence of $\tau_H/\tau_\alpha$ is similar to that 
of $\tau_F/\tau_\alpha$. This is not surprising because both $\tau_F$ and $\tau_H$ 
correspond to the time at which the distribution of particle displacements becomes
Gaussian. Note that $\tau_H$ is larger than $\tau_F$ for our choice of the cutoffs used for
measuring these time scales. The values of the ratio $\tau_H/\tau_\alpha$ lie in the range
$\sim 6-35$  for temperatures in the range 
$T\in [1.000-0.450]$.

As suggested in Ref.\cite{Szamel2006}, the time scales $\tau_F$ and $\tau_H$
may be thought of as lower bounds of the lifetime of 
dynamic heterogeneity.
As reported in Refs.\cite{SKJCP2014,BDKJstat2016}, the distribution of 
diffusion constants is a good measure of dynamic heterogeneity 
in the system. The distribution of diffusion constants is obtained from 
the self part of the van Hove function using Lucy iteration~\cite{LUCY1974}.
We have followed the method described in \cite{SKJCP2014,BDKJstat2016}. 
In the top and bottom panels of Fig.~\ref{Diff_dist}, we show the
distribution of diffusion constants for $T= 0.450$ and $T = 1.000$, 
respectively, for different times for the 3dKA model. For a high temperature {\it e.g.} $T = 1.000$, 
the distribution does not 
have multiple peaks but the broadness of the distribution reflects the
underlying heterogeneity. 
For $T= 0.450$, one can see 
that there are multiple peaks in the distributions even at time scales
longer than $\tau_F$. It approaches unimodal behavior at much longer 
time scales. So heterogeneity survives even after the system starts to 
follow Fick's law of diffusion, which is in agreement with the conclusion of
previous studies \cite{Szamel2006}.

%\textcolor{red}{
\section{Relation between Different Time Scales}
In the previous section, we found that the temperature dependence 
of the different relaxation times ($\tau_\alpha, \tau_D, \tau_F$ 
and $\tau_H$) can be well-fitted by the VFT form with very similar
values of the divergence temperature $T_{VFT}$. This suggests that the different
time scales are related to one another via power laws.
In Fig.\ref{crossPlot}, we have shown such a 
relationship between different time scales. The  
linear dependence in a log-log plot confirms that a power-law relation describes
the data quite well, especially at low temperatures. The exponent $m$ that describes
the power-law relation between one of the time scales of diffusion and the 
$\alpha$-relaxation time is $\simeq 0.83$ for $\tau_D$, implying that $\tau_D$ increases
more slowly that $\tau_\alpha$ with decreasing temperature. The values of $m$ for $\tau_F$
and $\tau_H$ obtained from the fits, $m\simeq 1.21$ and $m\simeq1.17$, respectively, are
close to one another. This is consistent with the observation, mentioned above, that
these two time scales have very similar temperature dependence. The larger than unity values
of $m$ for these time scales imply that their growth with decreasing temperature is 
faster than the growth of $\tau_\alpha$. Since $\tau_F$ and $\tau_H$ are believed to
provide lower bounds to the lifetime of dynamic heterogeneity, this result implies
that dynamic heterogenieties exist over time scales that are much longer than the $\alpha$-relaxation
time at low temperatures. The values of $m$ for $\tau_D$, $\tau_F$ and $\tau_H$ also imply that $\tau_F$ and $\tau_H$ increase faster than $\tau_D$ as the temperature is decreased,
indicating that at low temperatures, there is a large time interval over which the MSD is linear in time, but the distribution of particle displacements is not Gaussian. Yet another consequence of the observed power-law dependence of the diffusion-related time scales on $\tau_\alpha$ is that the transition temperature of mode-coupling theory obtained by fitting the temperature dependence of $\tau_D$, $\tau_F$ or $\tau_H$ to a power law would be the same as that obtained from a power-law fit to the temperature dependence of $\tau_\alpha$. However, the exponent of the power-law would depend on which time scale is considered. 
%}  

\begin{figure}[!htpb]
\begin{center}
\includegraphics[scale=0.39]{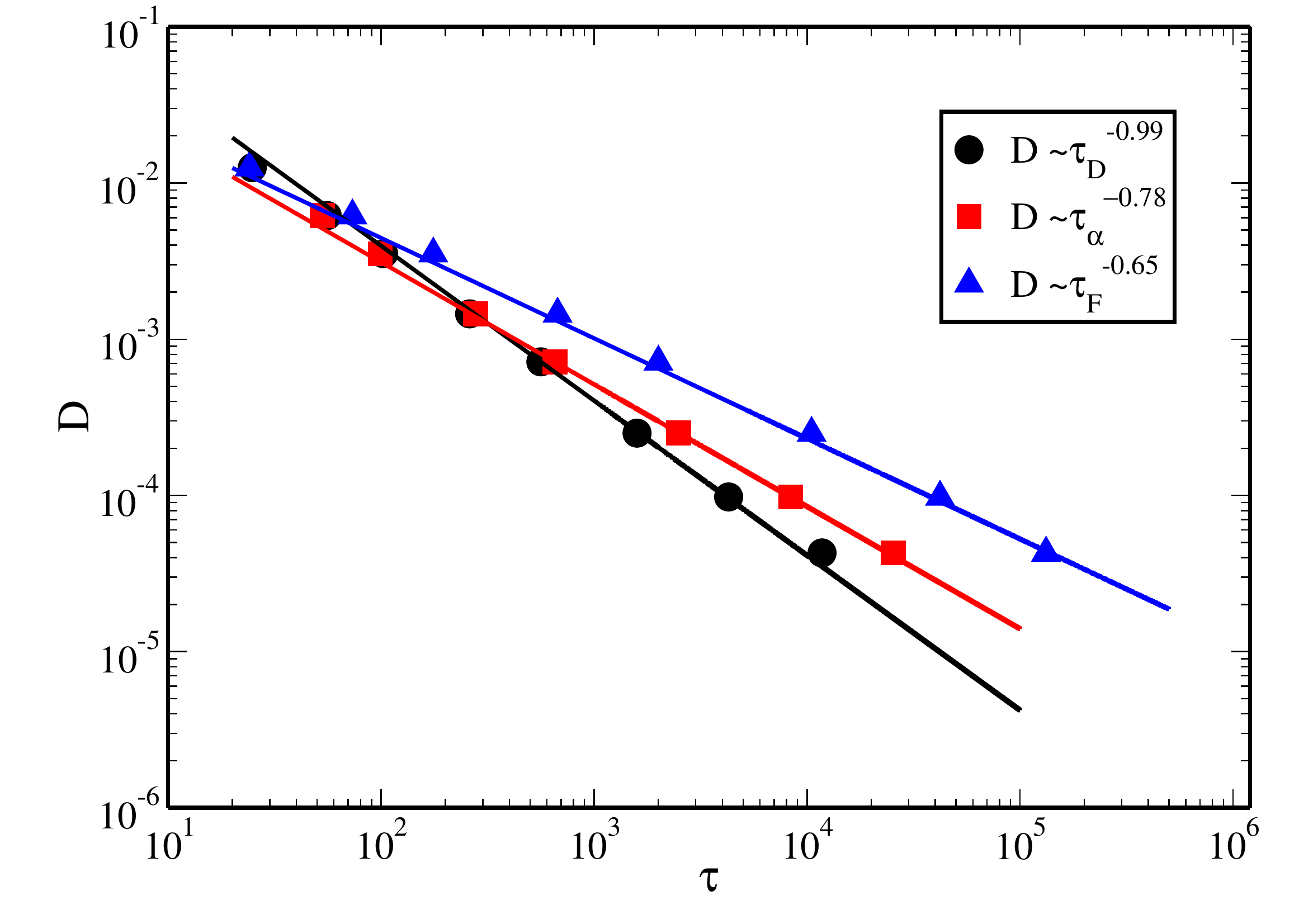}\quad
\includegraphics[scale=0.39]{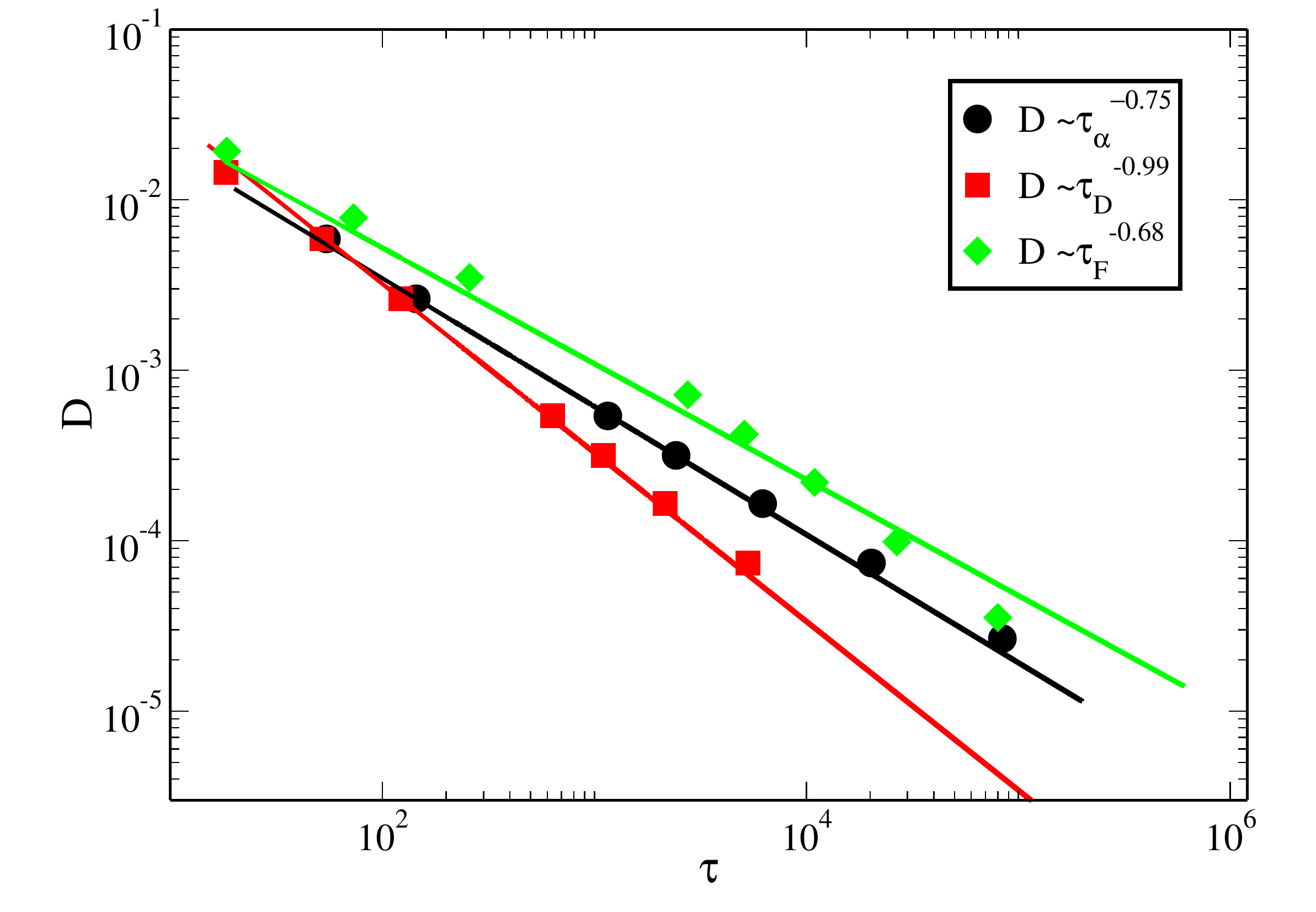}
\caption{{\bf Top panel:} The diffusion constant $D$ is plotted against different time scales for the 3dKA model system. The lines passing through the data points are power-law fits. {\bf Bottom panel:} Similar plot for the 3dR10 model.}
\label{SEviolation}
\end{center}
\end{figure}

\section{Stokes-Einstein relation and kinetic fragility}

The violation of the Stokes-Einstien ~\cite{Hansen1986,Einstein,LANDAU1987227}relation between the diffusion coefficient $D$ and the viscosity (or a suitable time scale $\tau$) in deeply supercooled liquids\cite{SKJCP2014,BDKJstat2016} is believed to be closely related to dynamic heterogeneity. It is interesting to examine the extent of violation of the SE relation when one of the time scales studied above is considered to be the relevant time scale $\tau$. It is known from earlier work~\cite{SE} that $D \propto 1/\tau_\alpha$ (i.e. the SE relation is satisfied) at high temperatures, but a modified relation, $D \propto \tau_\alpha^{-1+\omega}$, where $\omega$ is called the fractional SE exponent, is satisfied at lower temperatures.  The value of the 
exponent $(1-\omega)$ is $0.78$ and $0.75$  for 
3dKA and 3dR10 models respectively.

\begin{figure*}[htpb]
\begin{center}
\includegraphics[scale=0.35]{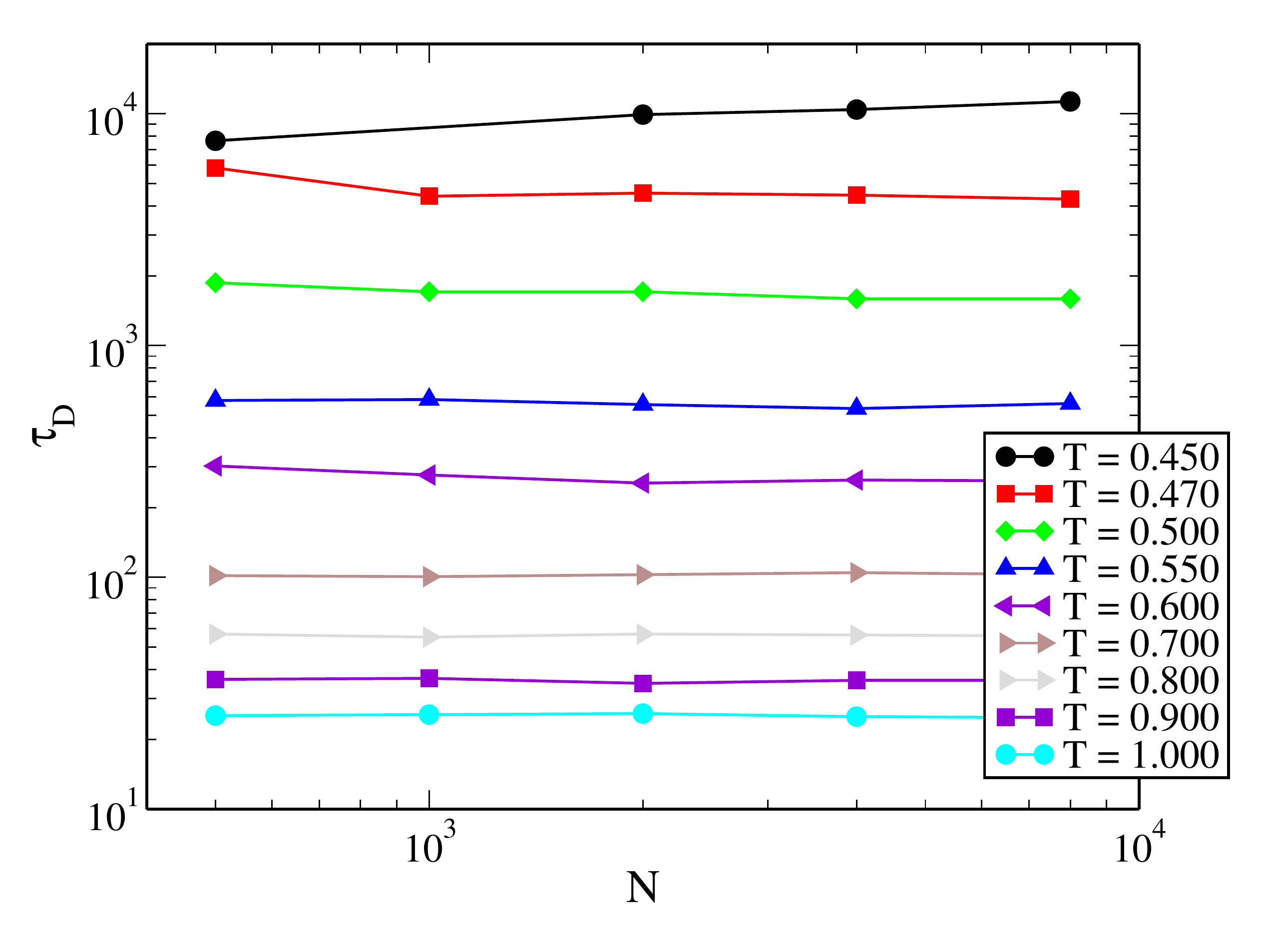}
\includegraphics[scale=0.31]{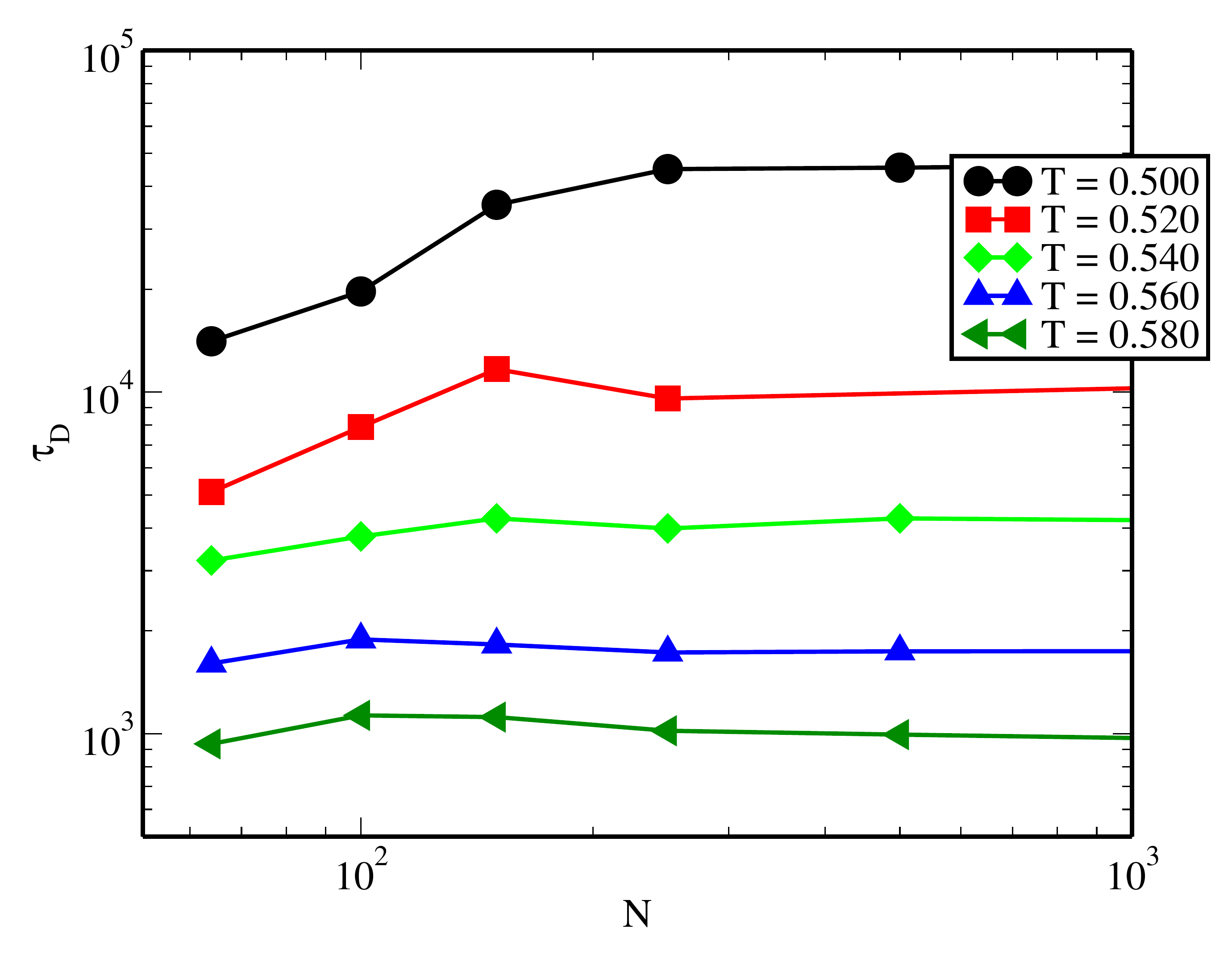}
\caption{Left Panel: System size dependence of $\tau_D$ at different 
temperatures for the 3dKA model system. The observed system size 
dependence for $N\in [500 - 10000]$ is very weak and only small 
dependence is observed at the low temperature. Right Panel: Zoomed 
in version of similar plot but for small system sizes 
$N\in [100 - 1000]$ for the lowest four temperatures for 3dR10 model.
Note for 3dR10 model, one can simulate smaller sizes as the range of the 
inter particle potential is small. See text for details}
\label{SystemSizeTau}
\end{center}
\end{figure*}
The power-law dependence of the time scales of diffusion on $\tau_\alpha$ implies that these time scales also satisfy a modified SE relation with values of the fractional SE exponent given by $(1-\omega)/m$. Since $m$ for $\tau_D$ for the 3dKA model is close to the value of $(1-\omega)$, the SE relation
is expected to be valid  for this time scale (i.e. $D \propto 1/\tau_D$) at low temperatures. On the other hand, the fractional SE exponent is expected to be small (close to $0.65$) for the two other time scales. The plots of the diffusion coefficient versus different time scales shown in Fig~\ref{SEviolation} illustrate this behavior. The SE relation is found to be satisfied for $\tau_D$, whereas the fractional SE exponent for $\tau_F$ is $0.65$. Similar behavior is found for the 3dR10 model, as illustrated in the bottom panel of Fig~\ref{SEviolation}.

As noted above, $\tau_\alpha$ satisfies the SE relation at temperatures higher than the onset temperature. Since the temperature dependence of the diffusion-related time scales is different from that of $\tau_\alpha$ even at these higher temperatures, these time scales do not satisfy the SE relation at temperatures near the onset temperature. The SE relation may be restored for these time scales at temperatures substantially higher than the onset temperature.

The power-law dependence of the diffusion-related time scales on $\tau_\alpha$ also implies
that the kinetic fragility parameter $\kappa$, obtained from a fit of the temperature dependence of a time scale $\tau$ to the VFT form, $\tau(T) = \tau_0 \exp[1/(\kappa(T/T_{VFT}-1)
]$, for these time scales is different from that for $\tau_\alpha$. The value of $\kappa$ for one of these time scales would be $1/m$ times the value for $\tau_\alpha$, implying that the fragility for $\tau_D$ is higher than that for $\tau_\alpha$, whereas the temperature dependence of $\tau_F$ and $\tau_H$ corresponds to a smaller value of the fragility.

\section{System-size dependence of time scales}
In Fig.~\ref{SystemSizeTau}, we show the system-size 
dependence of $\tau_{D}$ (left panel) at different 
temperatures for the 3dKA model. Almost no system-size 
dependence is observed at the higher temperatures, and only
a small dependence is found at the lowest temperature for system
sizes in the range $N\in [500 - 10000]$. We have done
simulations for smaller system sizes in the range 
$N\in [100 - 1000]$ for the 3dR10 model. Note that for this
model, one can simulate systems with smaller sizes as
the inter-particle potential has shorter range. The results are shown in the 
right panel of Fig.~\ref{SystemSizeTau} where one can 
clearly see the evidence for an initial increase of the time 
scale with system size at low temperatures.  
Note that $\tau_\alpha$ has a very different system size dependence 
- its value decreases with increasing system size at low temperatures\cite{karmakar2009growing}. 
This observation is puzzling and warrants further investigation. 
 
\section{Summary and Conclusions}
%\textcolor{red}{
In summary, we have estimated different time scales related to diffusion for two well-known
glass-forming liquids. One of these time scales, $\tau_D$, is the time at which the dependence
of the MSD on time becomes linear. The other, $\tau_F$ or $\tau_H$, represents the time at which the distribution of particle displacements becomes Gaussian. The actual values of these
time scales depend on the cutoff or tolerance factor used in their measurement from MD simulations. However, our investigation of the dependence of the time scales on the cutoff suggests that the qualitative behavior we find is independent of the choice of the cutoff. The results obtained for the two model systems are similar.
 
We show that both $\tau_D$ and $\tau_F$ are larger than the $\alpha$-relaxation time $\tau_\alpha$ in the temperature range considered in our simulations. At low temperatures ($T$ lower than the onset temperature $T_{on}$), the growth of $\tau_D$ with decreasing temperature is slower, whereas the growth of $\tau_F$ is faster than that of $\tau_\alpha$. Between the two diffusion-related time scales, $\tau_F$ is substantially larger than $\tau_D$, indicating that there is a long time interval over which the MSD is linear in time, but the distribution of the displacements is not Gaussian. If both linear dependence of the MSD on time and Guassian distribution of displacements are considered to be essential features of Fickian diffusion, then the longer time scale $\tau_F$ should be identified as the time at which Fickian diffusion sets in. Our results then imply that Fickian diffusion sets in at times much longer than the $\alpha$-relaxation time in deeply supercooled liquids. This is in agreement with earlier results~\cite{Szamel2006}.

We find that the ratio $\tau_D/\tau_\alpha$ is a non-monotonic function of temperature and it peaks at a temperature close to the onset temperature $T_{on}$ at which the dynamics starts being influenced by the energy landscape. This observation suggests an intriguing connection between the behavior of the MSD as a function of time and the influence of the energy landscape on the dynamics. Another interesting observation is that $\tau_D$ is close to the bond-breakage time scale $\tau_{BB}$ for temperatures lower than $T_{on}$.  This result raises questions that merit further investigation. 

The diffusion-related time scales $\tau_D$ and $\tau_F$ 
and the $\alpha$-relaxation time $\tau_\alpha$ are mutually correlated and
the relation between pairs of them is well-described by a power law at low temperatures.
This implies that  if any one of them diverges at some  
temperature, then the others also diverge at the same temperature. Thus, the VFT temperature
and the critical temperature of mode-coupling theory are the same for all these time scales. The power-law relation also implies that the fractional SE exponent and the kinetic fragility associated with the diffusion-related time scales are simply related to those obtained for $\tau_\alpha$. An interesting result in this context is that the time scale $\tau_D$ and the diffusion coefficient $D$ satisfy the SE relation at temperatures lower than $T_{on}$, implying that the time at which the MSD begins to be linear in time and the value of the diffusion coefficient obtained from this linear dependence are strongly related to each other ($D\propto 1/\tau_D$). 

On the other hand, the
Fickian time scale $\tau_F$ exhibits strong violation on the SE relation, with a fractional SE exponent close to $0.65$. This is puzzling because  one would naively expect the SE
relation to be valid when the diffusion is Fickian. However, the violation of the SE relation is believed to be caused by dynamic heterogeneity and the lifetime of dynamic heterogeneity may be longer than $\tau_F$. We have found some evidence for this from a calculation of the distribution of local diffusion constants (Fig.\ref{Diff_dist}) which exhibits multiple peaks for time scales longer than $\tau_F$, indicating that the lifetime of the heterogeneity is longer than the Fickian time scale. This is consistent with the suggestion~\cite{Szamel2006} that $\tau_F$ is a lower bound to the lifetime of dynamic heterogeneity. 

All these observations, raises several interesting 
questions about characteristics of the dynamics of supercooled 
liquids. Answers to these questions will be important in 
understanding the role of dynamic heterogeneity in glassy relaxation.

\acknowledgements{
This project was funded by intramural funds at TIFR Hyderabad from the 
Department of Atomic Energy (DAE).}
 
\bibliographystyle{ieeetr}
\bibliography{Fickian}

\begin{thebibliography}{10}

\bibitem{09Cav}
A.~Cavagna, ``Supercooled liquids for pedestrians,'' {\em Physics Reports},
  vol.~476, no.~4, pp.~51 -- 124, 2009.

\bibitem{11BB}
L.~Berthier and G.~Biroli, ``Theoretical perspective on the glass transition
  and amorphous materials,'' {\em Rev. Mod. Phys.}, vol.~83, pp.~587--645, Jun
  2011.

\bibitem{14KDS}
S.~Karmakar, C.~Dasgupta, and S.~Sastry, ``Growing length scales and their
  relation to timescales in glass-forming liquids,'' {\em Annual Review of
  Condensed Matter Physics}, vol.~5, no.~1, pp.~255--284, 2014.

\bibitem{05Berthier}
L.~Berthier, G.~Biroli, J.-P. Bouchaud, L.~Cipelletti, D.~E. Masri,
  D.~L{\textquoteright}H{\^o}te, F.~Ladieu, and M.~Pierno, ``Direct
  experimental evidence of a growing length scale accompanying the glass
  transition,'' {\em Science}, vol.~310, no.~5755, pp.~1797--1800, 2005.

\bibitem{06BBMR}
G.~Biroli, J.-P. Bouchaud, K.~Miyazaki, and D.~R. Reichman, ``Inhomogeneous
  mode-coupling theory and growing dynamic length in supercooled liquids,''
  {\em Phys. Rev. Lett.}, vol.~97, p.~195701, Nov 2006.

\bibitem{08BBCGV}
G.~Biroli, J.-P. Bouchaud, A.~Cavagna, T.~S. Grigera, and P.~Verrocchio,
  ``Thermodynamic signature of growing amorphous order in glass-forming
  liquids,'' {\em Nature Physics}, vol.~4, pp.~771 EP --, Aug 2008.

\bibitem{09KDS}
S.~Karmakar, C.~Dasgupta, and S.~Sastry, ``Growing length and time scales in
  glass-forming liquids,'' {\em Proceedings of the National Academy of
  Sciences}, vol.~106, no.~10, pp.~3675--3679, 2009.

\bibitem{15KDSRoPP}
S.~Karmakar, C.~Dasgupta, and S.~Sastry, ``Length scales in glass-forming
  liquids and related systems: a review,'' {\em Reports on Progress in
  Physics}, vol.~79, p.~016601, dec 2015.

\bibitem{RFOT}
T.~R. Kirkpatrick, D.~Thirumalai, and P.~G. Wolynes, ``Scaling concepts for the
  dynamics of viscous liquids near an ideal glassy state,'' {\em Phys. Rev. A},
  vol.~40, pp.~1045--1054, Jul 1989.

\bibitem{RFOT1}
V.~Lubchenko and P.~G. Wolynes, ``Theory of structural glasses and supercooled
  liquids,'' {\em Annual Review of Physical Chemistry}, vol.~58, no.~1,
  pp.~235--266, 2007.
\newblock PMID: 17067282.

\bibitem{RFOT2}
G.~Biroli and J.~Bouchaud, ``Structural glasses and supercooled liquids:
  Theory, experiment, and applications,'' 2012.

\bibitem{Ediger2000}
M.~D. Ediger, ``Spatially heterogeneous dynamics in supercooled liquids,'' {\em
  Annual Review of Physical Chemistry}, vol.~51, no.~1, pp.~99--128, 2000.
\newblock PMID: 11031277.

\bibitem{RankoRichart2002}
R.~Richert, ``Heterogeneous dynamics in liquids: fluctuations in space and
  time,'' {\em Journal of Physics: Condensed Matter}, vol.~14, no.~23, p.~R703,
  2002.

\bibitem{Andersen6686}
H.~C. Andersen, ``Molecular dynamics studies of heterogeneous dynamics and
  dynamic crossover in supercooled atomic liquids,'' {\em Proceedings of the
  National Academy of Sciences}, vol.~102, no.~19, pp.~6686--6691, 2005.

\bibitem{Szamel2006}
G.~Szamel and E.~Flenner, ``Time scale for the onset of fickian diffusion in
  supercooled liquids,'' {\em Phys. Rev. E}, vol.~73, p.~011504, Jan 2006.

\bibitem{Ediger1995}
M.~T. Cicerone and M.~D. Ediger, ``Relaxation of spatially heterogeneous
  dynamic domains in supercooled orthoterphenyl,'' {\em The Journal of Chemical
  Physics}, vol.~103, no.~13, pp.~5684--5692, 1995.

\bibitem{Ediger1999}
C.-Y. Wang and M.~D. Ediger, ``How long do regions of different dynamics
  persist in supercooled o-terphenyl?,'' {\em The Journal of Physical Chemistry
  B}, vol.~103, no.~20, pp.~4177--4184, 1999.

\bibitem{Deschenes255}
L.~A. Deschenes and D.~A. Vanden~Bout, ``Single-molecule studies of
  heterogeneous dynamics in polymer melts near the glass transition,'' {\em
  Science}, vol.~292, no.~5515, pp.~255--258, 2001.

\bibitem{Bohmer1996}
R.~B{\"o}hmer, G.~Hinze, G.~Diezemann, B.~Geil, and H.~Sillescu, ``Dynamic
  heterogeneity in supercooled ortho-terphenyl studied by multidimensional
  deuteron nmr,'' {\em EPL (Europhysics Letters)}, vol.~36, no.~1, p.~55, 1996.

\bibitem{Bohmer1998}
R.~B{\"o}hmer, G.~Diezemann, G.~Hinze, and H.~Sillescu, ``A nuclear magnetic
  resonance study of higher-order correlation functions in supercooled
  ortho-terphenyl,'' {\em The Journal of Chemical Physics}, vol.~108, no.~3,
  pp.~890--899, 1998.

\bibitem{Rajsekhar2018}
R.~Das, I.~Tah, and S.~Karmakar, ``Possible universal relation between short
  time $\beta$-relaxation and long time $\alpha$-relaxation in glass-forming
  liquids,'' {\em The Journal of Chemical Physics}, vol.~149, no.~2, p.~024501,
  2018.

\bibitem{KarmakarPrl2016}
S.~Karmakar, C.~Dasgupta, and S.~Sastry, ``Short-time beta relaxation in
  glass-forming liquids is cooperative in nature,'' {\em Phys. Rev. Lett.},
  vol.~116, p.~085701, Feb 2016.

\bibitem{Berthier1797}
L.~Berthier, G.~Biroli, J.-P. Bouchaud, L.~Cipelletti, D.~E. Masri,
  D.~L{\textquoteright}H{\^o}te, F.~Ladieu, and M.~Pierno, ``Direct
  experimental evidence of a growing length scale accompanying the glass
  transition,'' {\em Science}, vol.~310, no.~5755, pp.~1797--1800, 2005.

\bibitem{Szamel2010}
E.~Flenner and G.~Szamel, ``Dynamic heterogeneity in a glass forming fluid:
  Susceptibility, structure factor, and correlation length,'' {\em Phys. Rev.
  Lett.}, vol.~105, p.~217801, Nov 2010.

\bibitem{Lacevic2003}
N.~La{\v c}evi{\'c}, F.~W. Starr, T.~B. Schr{\o}der, and S.~C. Glotzer,
  ``Spatially heterogeneous dynamics investigated via a time-dependent
  four-point density correlation function,'' {\em The Journal of Chemical
  Physics}, vol.~119, no.~14, pp.~7372--7387, 2003.

\bibitem{PhysRevLett.104.165703}
C.~Crauste-Thibierge, C.~Brun, F.~Ladieu, D.~L'H\^ote, G.~Biroli, and J.-P.
  Bouchaud, ``Evidence of growing spatial correlations at the glass transition
  from nonlinear response experiments,'' {\em Phys. Rev. Lett.}, vol.~104,
  p.~165703, Apr 2010.

\bibitem{Hansen1986}
J.~P. Hansen and J.~R. McDonald, {\em Theory of Simple Liquids}.
\newblock Academic, London, 2~ed., 1986.

\bibitem{Einstein}
A.~Einstein, ``{\"U}ber die von der molekularkinetischen theorie der w{\"a}rme
  geforderte bewegung von in ruhenden fl{\"u}ssigkeiten suspendierten
  teilchen,'' {\em Annalen der Physik}, vol.~322, no.~8, pp.~549--560, 1905.

\bibitem{LANDAU1987227}
L.~LANDAU and E.~LIFSHITZ, ``Chapter vi - diffusion,'' in {\em Fluid Mechanics
  (Second Edition)} (L.~LANDAU and E.~LIFSHITZ, eds.), pp.~227 -- 237,
  Pergamon, second edition~ed., 1987.

\bibitem{Fulcher1925}
G.~S. Fulcher, ``Analysis of recent measurements of the viscosity of glasses,''
  {\em Journal of the American Ceramic Society}, vol.~8, no.~6, pp.~339--355,
  1925.

\bibitem{Vogel1921}
D.~H. Vogel., ``Das temperaturabhaengigkeitsgesetz der viskositaet von
  fluessigkeiten,'' {\em Physikalische Zeitschrift}, vol.~22, p.~645, 1921.

\bibitem{gotze1991liquids}
W.~G{\"o}tze, ``Liquids, freezing and the glass transition,'' 1991.

\bibitem{Gotze_1992}
W.~Gotze and L.~Sjogren, ``Relaxation processes in supercooled liquids,'' {\em
  Reports on Progress in Physics}, vol.~55, pp.~241--376, mar 1992.

\bibitem{95KA}
W.~Kob and H.~C. Andersen, ``Testing mode-coupling theory for a supercooled
  binary lennard-jones mixture i: The van hove correlation function,'' {\em
  Phys. Rev. E}, vol.~51, pp.~4626--4641, May 1995.

\bibitem{12KLP}
S.~Karmakar, E.~Lerner, and I.~Procaccia, ``Direct estimate of the static
  length-scale accompanying the glass transition,'' {\em Physica A: Statistical
  Mechanics and its Applications}, vol.~391, no.~4, pp.~1001 -- 1008, 2012.

\bibitem{vanHove1954}
L.~Van~Hove, ``Correlations in space and time and born approximation scattering
  in systems of interacting particles,'' {\em Phys. Rev.}, vol.~95,
  pp.~249--262, Jul 1954.

\bibitem{Shiba2016}
H.~Shiba, Y.~Yamada, T.~Kawasaki, and K.~Kim, ``Unveiling dimensionality
  dependence of glassy dynamics: 2d infinite fluctuation eclipses inherent
  structural relaxation,'' {\em Phys. Rev. Lett.}, vol.~117, p.~245701, Dec
  2016.

\bibitem{Yamamoto1997}
R.~Yamamoto and A.~Onuki, ``Kinetic heterogeneities in a highly supercooled
  liquid,'' {\em Journal of the Physical Society of Japan}, vol.~66, no.~9,
  pp.~2545--2548, 1997.

\bibitem{Indra2018}
I.~Tah, S.~Sengupta, S.~Sastry, C.~Dasgupta, and S.~Karmakar, ``Glass
  transition in supercooled liquids with medium-range crystalline order,'' {\em
  Phys. Rev. Lett.}, vol.~121, p.~085703, Aug 2018.

\bibitem{Sastry:1998aa}
S.~Sastry, P.~G. Debenedetti, and F.~H. Stillinger, ``Signatures of distinct
  dynamical regimes in the energy landscape of a glass-forming liquid,'' {\em
  Nature}, vol.~393, no.~6685, pp.~554--557, 1998.

\bibitem{Miyazaki_2011}
A.~Ikeda and K.~Miyazaki, ``Slow dynamics of the high density gaussian core
  model,'' {\em The Journal of Chemical Physics}, vol.~135, no.~5, p.~054901,
  2011.

\bibitem{SKJCP2014}
S.~Sengupta and S.~Karmakar, ``Distribution of diffusion constants and
  stokes-einstein violation in supercooled liquids,'' {\em The Journal of
  Chemical Physics}, vol.~140, no.~22, p.~224505, 2014.

\bibitem{BDKJstat2016}
B.~P. Bhowmik, R.~Das, and S.~Karmakar, ``Understanding the stokes--einstein
  relation in supercooled liquids using random pinning,'' {\em Journal of
  Statistical Mechanics: Theory and Experiment}, vol.~2016, no.~7, p.~074003,
  2016.

\bibitem{LUCY1974}
L.~B. {Lucy}, ``{An iterative technique for the rectification of observed
  distributions},'' {\em The Astronomical Journal}, vol.~79, p.~745, Jun 1974.

\bibitem{karmakar2009growing}
S.~Karmakar, C.~Dasgupta, and S.~Sastry, ``Growing length and time scales in
  glass-forming liquids,'' {\em Proceedings of the National Academy of
  Sciences}, vol.~106, no.~10, pp.~3675--3679, 2009.

\end{thebibliography}

\end{document}